\documentclass[sigconf]{acmart}
\usepackage{enumitem}
\usepackage{amsmath}
\usepackage{multirow}
\usepackage{bm}
\usepackage{graphicx}
\usepackage{threeparttable}
\usepackage[ruled, linesnumbered]{algorithm2e}
\usepackage{stfloats}
\usepackage{enumitem}
\setlist[itemize]{noitemsep, topsep=2pt}

\AtBeginDocument{%
  \providecommand\BibTeX{{%
    \normalfont B\kern-0.5em{\scshape i\kern-0.25em b}\kern-0.8em\TeX}}}

\copyrightyear{2022} 
\acmYear{2022} 
\setcopyright{acmcopyright}\acmConference[ICPC '22]{30th International Conference on Program Comprehension}{May 16--17, 2022}{Virtual Event, USA}
\acmBooktitle{30th International Conference on Program Comprehension (ICPC '22), May 16--17, 2022, Virtual Event, USA}
\acmPrice{15.00}
\acmDOI{10.1145/3524610.3527901}
\acmISBN{978-1-4503-9298-3/22/05}

\begin{document}

\title{HatCUP: Hybrid Analysis and Attention based Just-In-Time Comment Updating}


% \affiliation{%

\author{Hongquan Zhu}
\affiliation{%
  \institution{State Key Lab for Novel Software Technology, Nanjing University}
  \city{Nanjing}
  \country{China}
}
\email{hqzhu@smail.nju.edu.cn}

\author{Xincheng He}
\affiliation{%
  \institution{State Key Lab for Novel Software Technology, Nanjing University}
  \city{Nanjing}
  \country{China}
}
\email{xinchenghe2016@gmail.com}

\author{Lei Xu}
\authornote{Corresponding author.}
\affiliation{%
 \institution{State Key Lab for Novel Software Technology, Nanjing University}
 \city{Nanjing}
 \country{China}
}
\email{xlei@nju.edu.cn}

\renewcommand{\shortauthors}{Zhu, He and Xu}

\begin{abstract}
When changing code, developers sometimes neglect updating the related comments, bringing inconsistent or outdated comments. These comments increase the cost of program understanding and greatly reduce software maintainability. Researchers have put forward some solutions, such as CUP and HEBCUP, which update comments efficiently for simple code changes (i.e. modifying of a single token), but not good enough for complex ones. In this paper, we propose an approach named \textbf{HatCUP} (\textbf{\underline{H}}ybrid \textbf{\underline{A}}nalysis and A\textbf{\underline{t}}tention based \textbf{\underline{C}}omment \textbf{\underline{UP}}dater), to provide a new mechanism for comment updating task. HatCUP pays attention to hybrid analysis and information. First, HatCUP considers the code structure change information and introduces a structure-guided attention mechanism combined with code change graph analysis and optimistic data flow dependency analysis. With a generally popular RNN-based encoder-decoder architecture, HatCUP takes the action of the code edits, the syntax, semantics and structure code changes, and old comments as inputs and generates a structural representation of the changes in the current code snippet. Furthermore, instead of directly generating new comments, HatCUP proposes a new \emph{edit or non-edit} mechanism to mimic human editing behavior, by generating a sequence of edit actions and constructing a modified RNN model to integrate newly developed components. Evaluation on a popular dataset demonstrates that HatCUP outperforms the state-of-the-art deep learning-based approaches (CUP) by 53.8\% for accuracy, 31.3\% for recall and 14.3\% for METEOR of the original metrics. Compared with the heuristic-based approach (HEBCUP), HatCUP also shows better overall performance.
% Our idea is to imitate people's behaviour of modifying through insertion, deleting and updating, instead of generating a new comment from the beginning.

% We conducted several experiments to evaluate our approach on a popular dataset used in previous studies. Our results show that HatCUP improves relatively over the state-of-the-art deep learning-based approaches (CUP) in all the metrics. Compared with the heuristic-based approach (HEBCUP), HatCUP achieves better performance in complex scenario.

\end{abstract}






\maketitle

\section{Introduction}
As the complexity of software projects and the frequency of software product iterations continue to increase, program comprehension is becoming more important throughout the software development process. As recently shown by Xia et al. \cite{xia2017measuring}, 58\% of developers’ time was spent in comprehending code. In addition to the code itself, code comments are considered as the most important form of documentation for program comprehension \cite{de2005study}. Source code is constantly evolving, with developers regularly refactoring and integrating new functionality; however, code comments are often ignored when the code goes through changes \cite{ratol2017detecting,tan2007icomment,wen2019large}, leading to the inconsistency between code and comments that not only brings about confusion in software development and maintenance \cite{hu2018deep} but can also result in bugs \cite{tan2007icomment}. %To deal with this problem, we propose an approach that can give comment updating suggestions automatically when the associated code snippets are changed.

Comment generation aims to summarize code snippets with code representations \cite{ahmad2020transformer,hu2018deep,liang2018automatic,wan2018improving,zhang2020retrieval}  by generating an entirely new comment related to the current version of the code. However, they cannot retain some content intended to be highlighted in the existing comment. Recently, some approaches have been proposed to focus on automatic comment updating. For example, Liu et al. \cite{liu2020automating,liu2021just} propose a just-in-time technique, called CUP, to cope with the problem of the widespread presence of inconsistent comments. The core idea of CUP is to leverage a neural sequence-to-sequence model to learn comment update patterns from old comments and changed code tokens.

Although CUP has good performance on comment updating, it has some limitations. Lin et al. \cite{lin2021automated} find that since major correct comments generated are related to modifying a single token,  CUP tends to fail when its actual application scope is limited due to frequent updates. Hence, Lin et al. propose a heuristic-based comment updater HEBCUP that has the edge over CUP by focusing on the changed code patterns. However, HEBCUP also lacks efficiency faced with processing complex updating. When there are many changed code tokens or the code changes are not directly related to the old comment, CUP and HEBCUP may fail to make correct updates.

%In this paper, we try to deal with comment updating for complex scenarios. Therefore, we choose an RNN-based encoder-decoder architecture. Such a model has been shown to be effective for many Software Engineering (SE) tasks \cite{hu2018deep,liu2019automatic,tufano2019learning}. 

%To achieve automatic annotation updates that can cope with both simple and complex scenarios, our design strategies are (1) doing edits to the original comments wherever possible, rather than generating new comments, and (2) mining for structurally relevant information beyond code text changes.

This paper proposes a new approach called HatCUP to address comment updating in complex scenarios. Firstly, HatCUP considers more about the code structure change information. Instead of only focusing on code text changes, HatCUP pays attention to hybrid analysis and information with a structure-guided attention mechanism. It proposes a constraint-based optimistic data flow dependency analysis. Specifically, the derivation of such data-flow is not through conservative standard data-flow analysis, but rather similar to how humans derive data-flow, and more effective to the comment updating scenario.
Combined with constraint-based optimistic data flow dependency analysis and code change graph analysis, HatCUP can obtain the changed variable nodes and their associated dependencies on each other. Then, HatCUP constructs a change-guided and dependency-guided attention mechanism to help the model focus on changed syntax nodes and long-term dependencies among variables. Hence, HatCUP can collect complete information even when the code changes are not directly related to the old comment.

%For the second strategy, we introduce AST and structure-guided attention, which has not been used in previous work about this task. AST is an abstract representation of the syntactic structure of source code. It provides crucial code structure information for code understanding \cite{guo2021graphcodebert}. Of course, the introduction of AST does not imply the application of a graph neural network. Considering that our focus is on the code before and after the change, we need to find the connection between the two pieces of code. We obtain syntax change information with the state-of-the-art AST-diff tool GumTreeDiff \cite{falleri2014fine} by analyzing the AST of the two pieces of code separately. 

Secondly, with the core idea of imitating human editing behaviors, HatCUP considers more about edit actions(e.g., inserting, deleting and updating) on original comments instead of directly generating new comments. HatCUP modifies an RNN-based encoder-decoder model that is shown to be effective for many Software Engineering (SE) tasks \cite{hu2018deep,liu2019automatic,tufano2019learning} to integrate our newly developed component. Based on the model, HatCUP proposes a new \emph{edit or non-edit} mechanism to emphasize the possibility that a certain token in the old comment will be edited to match the new code patterns. Specifically, the \emph{edit or non-edit} mechanism leverages three different encoders to encode code changes, syntax changes and old comments. Then, HatCUP determines how the source code changes associated with the current decoding step changing the relevant parts of the old comment with three scenarios: (1) it decides whether a new edit action should be executed by generating an action-start keyword; (2) it preserves the current edit action by generating a common token; and (3) it suspends the current action until generating an action-end keyword. Finally, the decoder produces a series of edit actions, and HatCUP generates an updated comment based on the old comment and the corresponding edit actions.

To evaluate our approach, we use the same dataset in previous work\cite{liu2020automating,lin2021automated}, which contains code-comment co-change samples extracted from 1 496 popular engineered Java projects hosted on GitHub.

In summary, the contributions of this paper include:
\begin{itemize}[leftmargin=20pt]
\item \textbf{Hybrid analysis and attention}: 
We consider the code structure change information and introduce a structure-guided attention mechanism combined with code change graph analysis and optimistic data flow dependency analysis. Based on the multiple information about code changes(e.g.,  the action of the code edits, the syntax, semantics and structure code changes, and old comments), a structural representation of the changes in the current code snippet can be generated.
\item \textbf{A new mechanism for comment updating}: We propose a new mechanism, called \emph{edit or non-edit} mechanism. Instead of directly generating a new comment sentence from scratch, the \emph{edit or non-edit} mechanism generates a sequence of edit actions and constructs a modified RNN model to integrate newly developed component to mimic human editing behavior.
\item \textbf{Better performance}: HatCUP is shown to outperform the two state-of-the-art techniques and can reduce developers’ efforts in updating comments. The results show that HatCUP outperforms CUP by 53.8\% for accuracy, 31.3\% for recall and 14.3\% for METEOR of the original metrics. Compared with HEBCUP, HatCUP also shows better overall performance.
\item \textbf{Open Source}:We open source the replication package of our work, including the dataset, the source code of HatCUP, our trained model and test results. All data in the study are publicly available at GitHub$\footnote{https://github.com/HATCUP0/hatcup}$.
\end{itemize}

The rest of this paper is organized as follows: the motivating example is presented in Section \ref{sec:motivation}. The technical details of HatCUP are described in Section \ref{sec:approach}. The evaluation for our approach is shown in Section \ref{sec:evaluation}. Section \ref{sec:discussion} discusses the situations where our approach may fail and the threats to validity. Related work and conclusions are in Section \ref{sec:relatedwork} and Section \ref{sec:conclusion}.

\begin{figure*}[t]
  \centering
  \includegraphics[width=\linewidth]{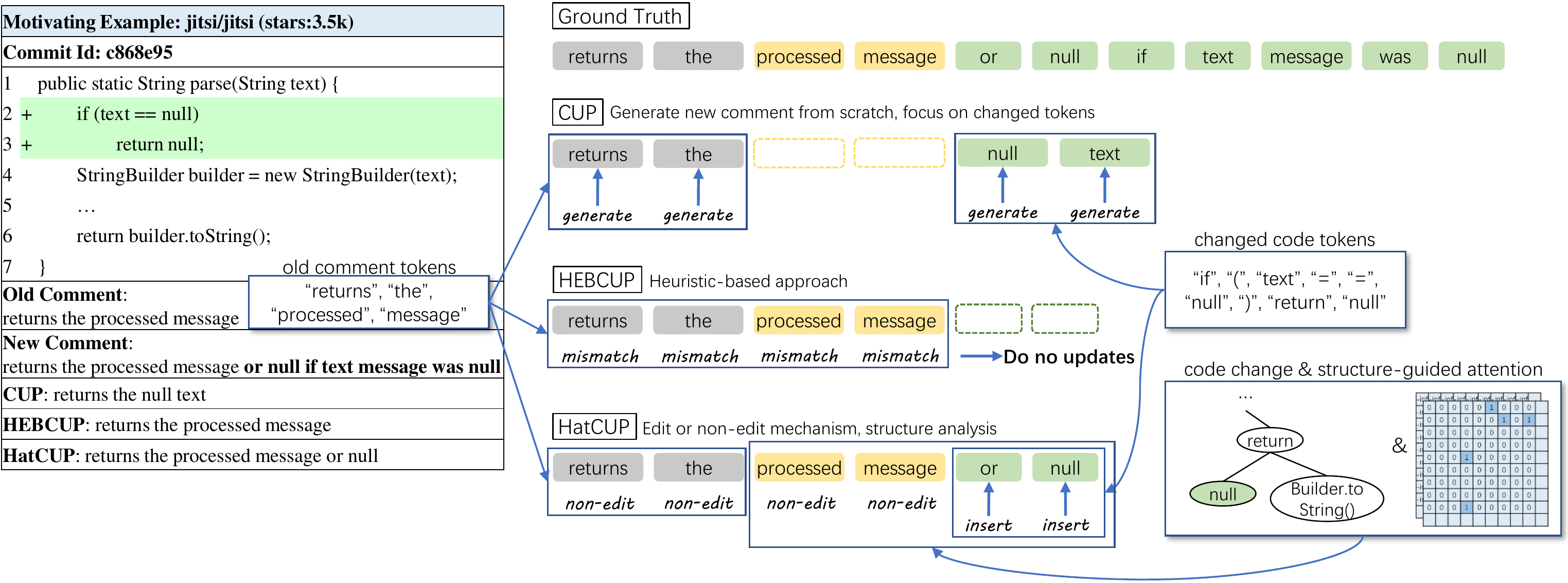}
  \caption{Motivating Example}
  \label{fig:motivation}
%   \vspace{-1em}  % 调整与下文的间距
\end{figure*}

\section{Motivation}
\label{sec:motivation}
% We show a motivating example from a popular GitHub repository to demonstrate why we propose HatCUP and how it works. We then present user scenarios of employing our proposed approach to address these problems.
% \subsection{Motivating Examples}

Some comment updating approaches based on neural model learning and heuristic rules are not sufficiently effective beyond simple updates \cite{lin2021automated}. We take CUP \cite{liu2020automating,liu2021just} and HEBCUP \cite{lin2021automated} as examples to demonstrate these limitations.

Figure \ref{fig:motivation} shows an example of stale comments we found in a real-world GitHub repository, \emph{Jitsi}. In an earlier version, a project developer added a judgment condition (lines 2-3) in the method \emph{parse()} to check if the variable \emph{text} is \emph{null}. However, the developer forgot to update the comment associated with this method, leading to a case of inconsistent comment. Fortunately, a developer found this problem and updated the comment later. 

To realize the goal of updating automatically, CUP leverages a neural sequence-to-sequence model to learn comment update patterns from old comments and changed code tokens, and generates a new comment \emph{"returns the null text"}.  However, since the return value of the target API \emph{parse()} has another value assignment related to the variable \emph{builder} in line 6 and the variable \emph{text} in lines 1-4, which is not included in the changed token sequence, the information of structure and data flow dependency in code is ignored. CUP only gives comments about \emph{null} and lacks the description \emph{"processed message"} related to the original return value in line 6.

As a heuristic-based approach, HEBCUP applies updates of the code (sub)tokens to the corresponding comment (sub)tokens by matching the (sub)token in the old comment with the (sub)token in the old code one by one. For instance, if a method name is changed from "getX" to "getY", its comment may be updated from “return X” to “return Y”. Its obvious disadvantage is that if the tokens of the code change do not match any tokens of the old comment, no updates can be made. Therefore, as shown in Figure \ref{fig:motivation}, HEBCUP could not add the new return value "null" to the comment. 

We propose HatCUP to provide a new mechanism for comment updating. In addition to the token sequences of code change and old comment, HatCUP considers more about the code structure change information and introduces a structure-guided attention mechanism combined with code change graph analysis and optimistic data flow dependency analysis. Hence, the addition of structure about return value \emph{null} and \emph{builder.toString()}, and the related data flow dependency about \emph{text} and \emph{builder} can be caught. Instead of generating new comments, we propose a new \emph{edit and non-edit} mechanism to capture the edit action and construct a modified RNN model to integrate newly developed components, with the core idea of mimicking human editing behaviors. Finally, the new comment about the \emph{insert} action \emph{"or null"} is updated successfully.

\section{Approach}
\label{sec:approach}
Our approach, HatCUP, consists of three phases: edit representation, model training, and comment updating. Specifically, for each code-comment co-change sample extracted from source code repositories, we first represent them as  edit sequences. Then, our model is trained using the preprocessed data. Finally, given code edits, syntax changes and associated the old comment, the trained model can automatically update the old comment to a new comment. In this section, we elaborate on the steps of our approach.

\subsection{Representing Edits}

In this phase, we convert code changes and comments into sequences. Different from CUP \cite{liu2020automating}, we adopt a new representation of code and comment changes. We will describe it in detail in 3.1.2.

\subsubsection{Data Pre-Processing}
In the preprocessing, each code snippet is split into tokens, and each identifier is tokenized based on camel casing and snake casing. For comments, HTML tags and comment symbols(e.g., "/*" and "//") are all filtered out. Then, each string will be tokenized by space. After that, compound words, which are the tokens constructed by concatenating multiple vocabulary words according to camel or snake conventions, are split into multiple tokens to reduce OOV (Out Of Vocabulary) words.

\subsubsection{Text Change Representation of Code and Comment}
After tokenization, the old and new code snippets are converted to two token sequences separately. We use difflib$\footnote{https://docs.python.org/3/library/difflib.html}$ to extract code edits. The code token sequence pair consists of a series of edit actions, which means editing an old code snippet to the new one.
We construct each edit action as \emph{<Action> [tokens] </Action>}, which has proven to be highly effective in other tasks in preliminary experiments \cite{shin2018towards}. We define four types of editing actions in our work: \emph{INSERT}, \emph{DEL}, \emph{UPDATE} and \emph{KEEP}. Notably, \emph{UPDATE} action must incorporate content both before and after the update, thus explicitly indicating which tokens in the old comment are to be replaced with the new tokens, so it has a slightly different structure:
\begin{center}
    \begin{tabular}{l}
    \emph{<UPDATEFROM> [old tokens]}\\
    \emph{<UPDATETO> [new tokens]}\\
    \emph{</UPDATE>}\\
    \end{tabular}
\end{center}

Especially, the comment edit representation is slightly different from the code representation. During inference, we only need to know the position and information of the changes made to the old comment. Therefore, we do not consider \emph{KEEP} type when building the sequence of edit actions, since we can copy tokens that are retained between the old and new comments, instead of generating them anew. For \emph{DEL} and \emph{UPDATE}, we can remove or replace exactly the corresponding content from the old comment. For \emph{INSERT} action, we design a method to determine which position of the old comment should be edited. Considering the example in Figure \ref{fig:motivation}, the raw sequence \emph{<INSERT>} \emph{or null ...} \emph{</INSERT>} does not contain information about where the new string should be inserted. Therefore, we select the minimum number of tokens before the insert position as a tag, so the place of insertion can be uniquely identified. Consequently, we will generate the edit action: \emph{<INSERTTAG>} \emph{message} \emph{<INSERT>} \emph{message or null if text message was null} \emph{</INSERT>}. Similar to the process of \emph{UPDATE}, this sequence indicates that “message” should be replaced with “message or null if text message was null,” effectively inserting “or null if text message was null” into the old comment.

\begin{figure*}[t]
  \centering
  \includegraphics[width=\linewidth]{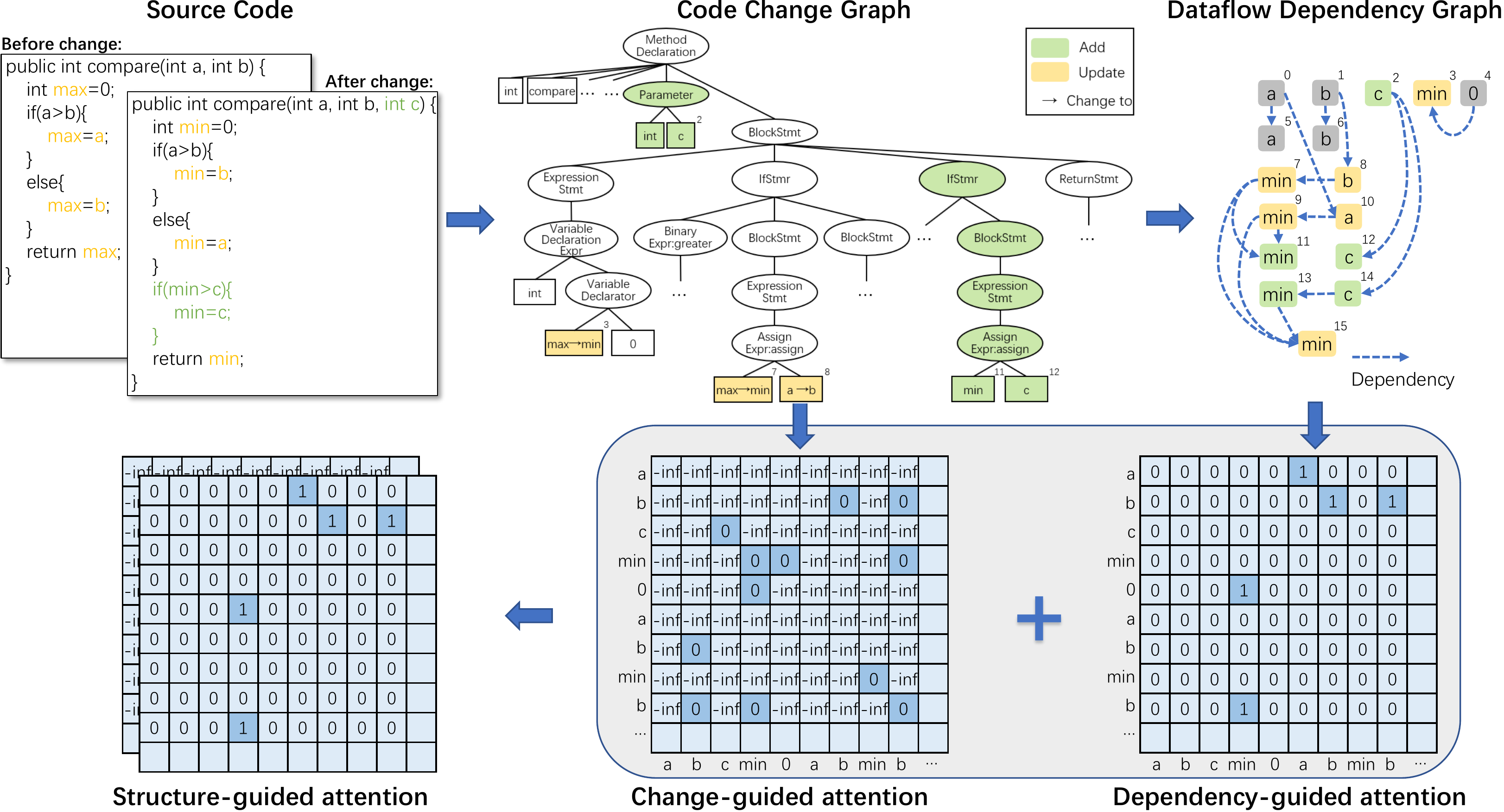}
  \caption{The structure-guided attention}
  \label{fig:attention}
  \vspace{-1em}  % 调整与下文的间距
\end{figure*}

\subsubsection{Syntax Change Representation}\label{section:ast_change}
AST (Abstract Syntax Tree) provides crucial structure information for code understanding \cite{guo2021graphcodebert}. Given a pair of old source code $C_1$ and new source code $C_2$, we can obtain syntax change information with GumTreeDiff \cite{falleri2014fine}. Variable nodes in the syntax tree will be used as the input of the syntax change encoder. However, this is not enough to help the model obtain structural information. We introduce a new attention mechanism for the comment updating scenario, the \emph{structure-guided attention} shown in Figure \ref{fig:attention}.

In Figure \ref{fig:attention}, a code change graph is created by analyzing the syntax change information provided by GumTreeDiff. Each node in this graph is a triple tuple $n_i$ = <$Operation, Type, Value$>. $Operation$ means the edit operation of a node from $C_1$ to $C_2$, namely: \emph{keep, insert, del, update}; $Type$ is the syntax type of a node, such as "SimpleName", "IfStatement", "Assignment" and so on; $Value$ is the value of a node (if the node has). 

To efficiently generate comment fragments for modified parts of the code, we propose change-guided attention. It focuses on the variable nodes in the code change graph involved in the change and their associated nodes. More formally, we introduce the following change-guided attention matrix:
\begin{equation}
    M_{i j}= \begin{cases}0 & \text { if }\left\langle n_i,n_j \right\rangle \in A \text{ and }\ n_i/n_j \in CN \\ $-inf$ & \text { otherwise }\end{cases}
\label{equa:M2}
\end{equation}
$A$ is the set of node pairs $\left\langle n_i,n_j \right\rangle$, in which $n_i$ and $n_j$ are the same node (i.e., $i=j$) or nodes of the same value, or there is an assignment relationship between them. $CN$ means the changed nodes set. After this attention matrix is passed to the softmax logical regression, all the parts set to \emph{-inf} will be ignored in the subsequent calculations. Consequently, the change-guided attention is designed to block out information other than changes.

\begin{table*}[t]
\caption{Constrained data flow extraction rules}
\label{tab:dataflow}
% \resizebox{\textwidth}{!}
\begin{threeparttable}          %这行要添加
{%
\begin{tabular}{c|l|l|l}
\hline
\textbf{Pattern name} & \textbf{Sub-pattern name} & \textbf{Flow derivation} & \textbf{Example} \\ \hline
\multirow{3}{*}{Output-flow} & $MethodParameter$ & $\left\{\forall e_{i} \in E, \forall u \in e_{i}\right\} \subseteq {OUT}(M)$ & $f(e_1,...,e_n)$ \\ \cline{2-4} 

                             & $PostfixExpression$ & $\left\{\forall e_{i} \in E, \forall u \in e_{i}\right\} \subseteq {OUT}(M)$ & $e$++ \\ \cline{2-4} 
                             
                             & $Assignment$(left-hand) & $\left\{\forall v \in O(l)\right\} \subseteq OUT(M)$ & $\bm{v}=e$ \\ \hline
                             
\multirow{7}{*}{Input-flow}  & $PrefixExpression$    & $\left\{\forall e_{i} \in E, \forall u \in e_{i}\right\} \subseteq {IN}(M)$   & $!e$   \\ \cline{2-4} 

                             & $InfixExpression$    & $\left\{\forall e_{i} \in E, \forall u \in e_{i}\right\} \subseteq {IN}(M)$   & $e_1<e_2$   \\ \cline{2-4} 
                             
                             & $PostfixExpression$    & $\left\{\forall e_{i} \in E, \forall u \in e_{i}\right\} \subseteq {IN}(M)$   & $e$++   \\ \cline{2-4}
                             
                             & $ContainerAccess$    & $\left\{\forall v \in C, \forall u \in e\right\} \subseteq {IN}(M)$   & $v[e]$   \\ \cline{2-4} 
                             
                             & $MethodInvocation$    & $\left\{\forall v \in E, \forall e_{i} \in E, \forall u \in e_{i}\right\} \subseteq {IN}(M)$  & $v.f(e_1,...,e_n)$   \\ \cline{2-4} 
                             
                             & $ReturnStatement$    & $\left\{\forall e_{i} \in E, \forall u \in e_{i}\right\} \subseteq {IN}(M)$   & $return$ $e_1,...,e_n$ \\ \cline{2-4} 
                             
                             & $Assignment$(right-hand)    & $\left\{\forall e \in O(r), \forall u \in e\right\} \subseteq IN(M)$   & $v=\bm{e}$   \\ \hline
\multirow{1}{*}{Preservation} & -        & $\left\{\forall e_i \in kill(E), \forall u \in e_i\right\} \nsubseteq IN(M) \cup OUT(M)$     & -  \\ \hline

\end{tabular}%
}
 \begin{tablenotes}
    \footnotesize 
    \item[1] $e \in E$ represents an expression; $\{u, v\} \subseteq e$ represents a variable object; $O(l)$ and $O(r)$ represent the left-hand and right-hand operand; $M$ is a method; $OUT(M)$ contains all the variable objects which has an output-flow.
 \end{tablenotes}
\end{threeparttable}
\end{table*}

We also constructed the data flow dependency graph. 
On the one hand, the data flow contains semantic code information, which is crucial for code understanding. 
%Taking $n=max\_num - min\_num$ as an example, developers may not follow the naming conventions, which will lead to confusion in understanding the semantics of the variable. The data flow provides a way to understand the semantic information of the variable $n$ to some degree, i.e., the value of n comes from $max\_num$ and $min\_num$ in the data flow. 
On the other hand, the data flow supports the model in considering long-term dependencies induced by using the same variables in distant locations. For example, there are six variables with the same name (i.e. $min^4$, $min^8$, $min^{10}$, $min^{12}$, $min^{14}$ and $min^{16}$) but different semantics in the data flow dependency graph in Figure \ref{fig:attention}. The graph demonstrates dependencies between variables and supports $min^{16}$ in paying more attention to $min^8$, $min^{10}$ and $min^{14}$ instead of $min^{4}$.

We determine the data flow dependencies based on each variable object's Output-flow and Input-flow. Inspired by PyART \cite{he2021pyart}, we define a set of constraint-based optimistic data flow dependency extraction rules according to the target of our task. Such data flow is neither sound nor complete, and it just appears to be concise and largely precise, thus facilitating our incorporation of this information in the model. As shown in Table \ref{tab:dataflow}, we have summarized 11 output or input-flow patterns, including: \emph{MethodParameter, PostfixExpression, Assignment(left-hand), PrefixExpression, InfixExpression, PostfixExpression, ContainerAccess, MethodInvocation, ReturnStatement and Assignment(right-hand)} rules. We filtered out some patterns defined as $Preservation$, because they are not common and therefore cannot provide enough data to support the learning of the model (and hence should be killed). We construct data flow dependencies from $OUT(M)$ collection to $IN(M)$ collection. For each variable $v$ in $IN(M)$, there must be a flow from the nearest variable $v$ in $OUT(M)$ before it. Taking the method in Figure \ref{fig:attention} as an example, $a^5$ (in $a>b$, InfixExpression) belongs to $IN(M)$, the nearest $a \in OUT(M)$ before $a^5$ is $a^0$ (in $compare(int\ a,...)$, $MethodParameter$). Consequently, there is a data flow from $a^0$ to $a^5$. In addition, there must be a flow from the right-hand operand to the left-hand operand in an $Assignment$ expression.

To represent the dependency relationship, we take a direct edge $\varepsilon$ = $\langle n_i, n_j\rangle$ from $n_i$ to $n_j$, which means that the value of the $j$-th node comes from the $i$-th node. The following dependency-guided attention matrix represents the dependency relationship:
\begin{equation}
    M_{i j}= \begin{cases}1 & \text { if } \left\langle n_{i}, n_{j}\right\rangle \in E \\ 0 & \text { otherwise }\end{cases}
\label{equa:M}
\end{equation}
$E$ is the set of directed edges, $\{\varepsilon_1, \varepsilon_2,...,\varepsilon_n\}$. Similar to the change-guided attention matrix, the dependency-guided attention matrix will also be passed to the softmax logical regression; after calculation, the parts set to 1 will have a higher attention score than the parts set to 0, which means that the model will pay more attention to the dependency relationship among the variables.

Finally, we compute the weighted sum of change-guided attention and relation-guided attention matrices to obtain the final structure-guided attention matrix. We filter the variable nodes associated with the change for the task of comment updating and try to preserve the dependencies among the nodes. In summary, we do not serialize the traversal of AST nodes as previous work and input the flattened sequence, which results in corrupted structural information.

\begin{figure*}[tp]
  \centering
  \includegraphics[width=\linewidth]{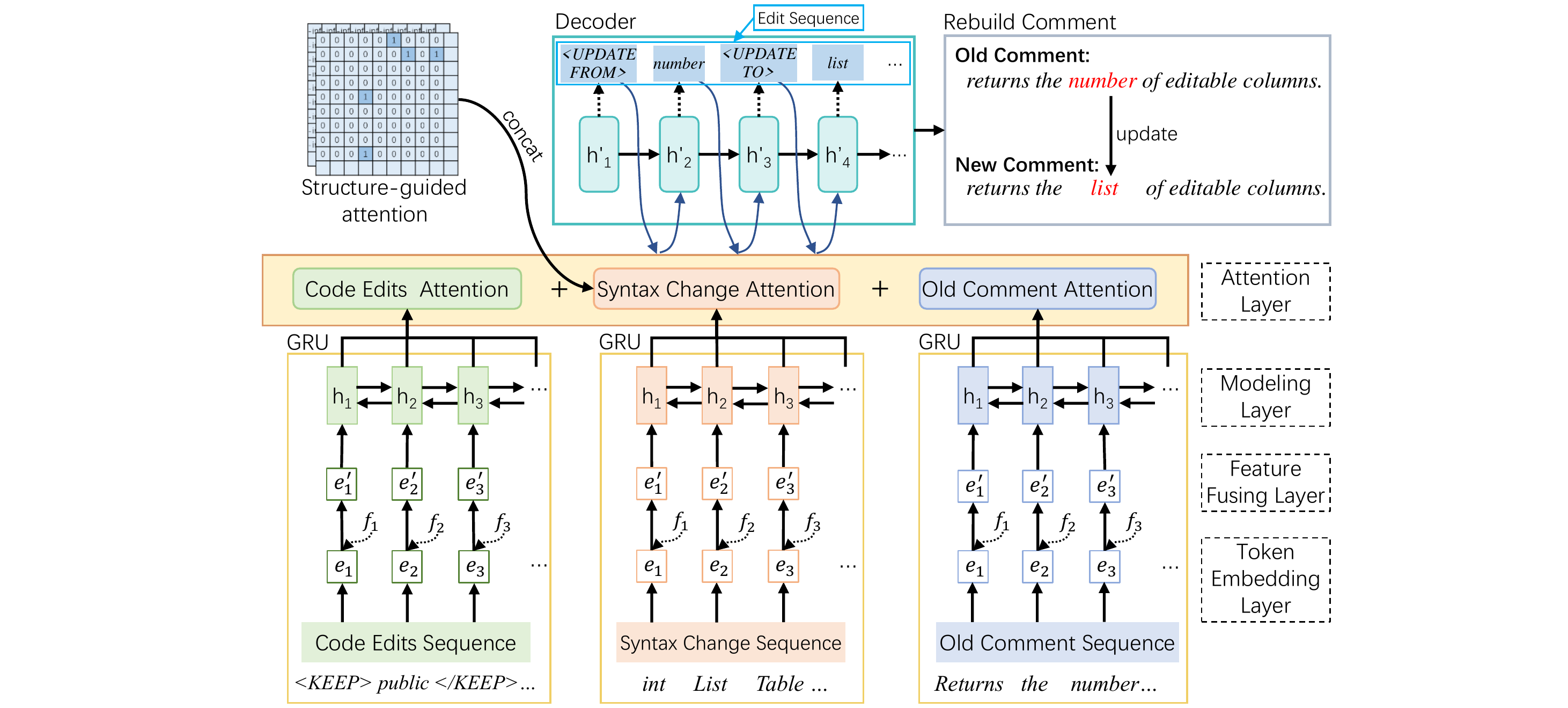}
  \caption{Architecture of HatCUP}
  \label{architecture}
  %\vspace{-1em}  % 调整与下文的间距
\end{figure*}

\subsection{Overview of Our Model}
Figure \ref{architecture} is the architecture of our model. Our model leverages three Bi-GRU (Bi-Directional Gated Recurrent Unit) encoders, i.e., \emph{Code Edits Encoder}, \emph{Syntax Change Encoder} and \emph{Old Comment Encoder}, and then generates a sequence of edit actions for the old comment through a GRU decoder. An encoder-side co-attention mechanism is leveraged to learn the relationships among code text change, code structure change and old comment. After the generation of the edit sequence, a parser will apply the edit actions to the old comment to obtain the updated comment.

\subsection{Encoders}
After extracting as features the sequences of sub-tokens and nodes for the codes, we need to convert those sequences into vector representations for the models in the later steps. In detail, HatCUP leverages three different encoders, to encode code edit sequences, syntax change sequences and old comments, respectively. We use multiple GRUs for the different structures and types of information. Multiple GRUs also help reduce the cross-influence between different contexts.

\subsubsection{Token Embedding Layer}
This layer is designed to map the various tokens, i.e., code tokens, comment tokens, edit action tokens and variable node tokens, into embeddings. We first created a vocabulary separately for code, variable nodes and comments. We choose to train the embedding layers from scratch instead of using a pretrained model, since the pretrained model may not contain some tokens, which will result in low effectiveness. Especially, for syntax change, we put the values of variable nodes as tokens into the encoder; 
% we replace the node value with <null> if it is empty. Note that we do not consider the order, as our aim is to capture dependencies between variables through the association of AST nodes. 
as for old comments, we input only the split subtokens.

\subsubsection{Feature Fusing Layer}
After the initial embedding of the tokens, we select extra features for each token, which has proven to be effective in learning associations between source code entities and comments \cite{panthaplackel2020associating}. These features are represented as one-hot vectors and are concatenated to code edit token embeddings or comment token embeddings.

For code edit sequences, which contain edit keywords, Java keywords, operators, variable names, etc., we need to make the model distinguish these tokens. If a token is not an edit keyword, we have indicator features for whether it is part of an \emph{INSERT}, \emph{DEL}, \emph{UPDATEFROM}, \emph{UPDATETO}, or \emph{KEEP} span. These features are particularly helpful for longer spans, as the edit keyword only appears at the beginning or end of the span. In addition, the tokens in return statements usually appear in comments, and we introduce these features to guide the model in identifying relevant tokens in the code edits sequence and the old comment sequence.

For syntax change, we also have indicator features such as that in code edits tokens, telling the model which operation a node belongs to, i.e., \emph{keep, insert, del, update}. Additionally, we take \emph{Type} into consideration. According to a large-scale empirical study conducted by Wen et al. \cite{wen2019large}, change types \emph{Variable Declaration} and \emph{Selection} are among those more likely to trigger comment updates, at the method and class level. These changes could severely impact the application logic (selection) or the data manipulated in the code \emph{Variable Declaration}.

Regarding old comments, we include whether a token matches an AST node that is \emph{insert, del, update} in AST-diff. This treatment helps align parts of old comments with AST changes, helping the model determine where the edits should be made.

\subsubsection{Modeling Layer}
This layer produces the hidden states of each token based on its contextual vector. For each context, we encode it with a GRU. In our approach, the three GRUs share a similar structure.

As shown in Figure \ref{architecture}, the input for each GRU is the sequence $V$ of $n$ vectors representing a context. Each vector represents a sub-token combining other features. For example, one GRU is used to obtain the hidden states of code edits contextual vectors. For each time step $t$, we input one vector $V_t$ in these $n$ vectors, and the GRU returns one hidden state vector $h_t$ as the output for this time step. By collecting all outputs for each time step, we have a sequence of hidden state vectors $H = [h_1, h_2, ..., h_n]$, which is the output of the GRU. The whole procedure can be expressed by Formula \ref{equal_h}. The other two GRUs share the same process.
\begin{equation}
  h_t = f(v_t, h_{t-1})
\label{equal_h}
\end{equation}

\subsection{Attention Layer}
In the modeling layer, we obtain the hidden states $H$ of the three GRUs separately. However, code edit, syntax change and old comment are represented independently. We cannot directly use them as input to the decoder to generate the result sequence. It is necessary to link and fuse their information to capture the relationships between different contexts. Therefore, we design three attention layers.

For each encoder, an attention layer is obtained by weighting and summing the outputs of all its timings. This attention layer contains information about the weight of each timing output, which is equivalent to identifying which text is important for the current token in the decoder. For instance, \emph{Code Edit Attention} is used to identify the parts of the code relevant to the target edit sequence to be generated; \emph{Old Comment Attention} is used to identify the notes needing to be edited in old comments. \emph{Syntax Change Encoder} will be slightly different, and it uses the structure-guided attention we defined in Section \ref{section:ast_change}.

For convenience, we take \emph{Code Edit Attention} as an example. The attention layer takes as input the code edits contextual vectors, i.e., $H = [h_1, h_2, ..., h_n]$, and outputs an attention-aware contextual vector $C=[c_1, c_2, ..., c_n]$ for each edit token in \emph{Code Edits Attention}. $c_t$ in $C$ is calculated as the weighted sum of the encoder’s hidden states:
\begin{equation}
    c_{t}=\sum_{i=1}^{n} \alpha_{ti} h_{i}
\end{equation}
\begin{equation}
    \alpha_{t i}=\frac{e^{r\left(h_{t-1}^{\prime}, h_{i}\right)}}{\sum_{i^{\prime} \neq i}^{n} e^{r\left(h_{t-1}^{\prime}, h_{i^{\prime}}\right)}}
\end{equation}
where $h'_{t-1}$ is the previous hidden state in the decoder, and $r$ is the function used to represent the strength for attention, approximated by a multi-layer neural network.

\subsection{Decoder}
By combining all the contextual vector outputs $C$ of all attention layers, we obtain a joint context; thus, the corresponding content from three input sequences is merged. We use a GRU as the decoder to generate a series of edit actions. At every decoding step, the previous hidden state $h'_{t-1}$ is used as the input for the attention layer and the output of the attention layer will be used as the input of the GRU at time step $t$. Therefore, the resulting vector contains information related to the current decoder state together with knowledge aggregated from relevant parts of code edits, AST-diffs and old comments.

Different from the previous work, we do not generate a full new comment. Instead, we generate a series of edit actions to show how to update the old comment. Specifically, the decoder must determine how the source code changes associated with the current decoding step will change the relevant parts of the old comment. At each step, the decoder decides whether a new edit action should be executed by generating an action-start keyword from \emph{INSERT}, \emph{DEL} or \emph{UPDATE} and continues the current edit action by generating a comment token; it will not stop the current action until an action-end keyword is generated. Since \emph{DEL} will include tokens in the old comment, and \emph{INSERT} tends to include tokens in the code, we add a pointer network to the decoder \cite{vinyals2015pointer} to accommodate copying tokens from code and comment. The decoder generates a series of edit actions. Consequently, we can generate an updated comment by parsing the old comment and the corresponding edit actions.

\subsection{Parsing Edit Sequences}
Since the decoder gives us a series of edit actions, we should align it with the old comment and apply it to obtain the updated comment. We denote the old comment as $S_{old}$, the predicted edit actions as $S_{edit}$ and the corresponding parsed output as $S_{new}$. This procedure involves simultaneously following pointers, from left to right, on $S_{old}$ and $S_{edit}$, which we refer to as $P_{old}$ and $P_{edit}$ respectively. As $P_{old}$ moves forward, the current token is copied into $S_{new}$ at each point, until the pointer reaches an edit location. Then $P_{edit}$ applies the edit action of the current position, and the span tokens corresponding to the action are copied into $S_{new}$ if applicable. Finally, $P_{edit}$ moves to the next action; so do cases involving deletions and replacements; $P_{old}$ is also advanced to the appropriate position. This process will repeat until the two pointers reach the end of their respective sequences.

\section{Evaluation}
\label{sec:evaluation}
\subsection{Dataset}
\label{sec:dataset}
We use the same dataset in HEBCUP \cite{lin2021automated} and CUP \cite{liu2020automating}. The authors of the two works built a dataset from 1,496 Java projects hosted on GitHub and design rules to automatically filter out some types of syntactic optimizations (i.e., the old and new comments are of the same meaning) which may introduce bias. The cleaned dataset finally contains 80,591, 8,827, and 9,204 method-comment co-change samples for training, validation, and test sets, discarding 6,183 instances in total.

\subsection{Research Questions}
To evaluate HatCUP, we propose the following research questions.

\textbf{RQ1: }How effective is HatCUP compared with the two state-of-the-art approaches, CUP and HEBCUP?

\textbf{RQ2: }How do the key components of HatCUP affect the result?
% What impact do the \emph{edit or non-edit} mechanism and  have when updating the comment?

\textbf{RQ3: }How effective is HatCUP when dealing with complex scenarios?

% \subsection{Baselines}
% TODO

\subsection{Experiment Setup}
We conducted our experiment on Ubuntu 18.04.6 with Intel(R) Xeon(R) Gold 5118 CPU @ 2.30GHz. We utilized 1 NVIDIA Tesla V100 GPU to train and evaluate our model. The model was implemented in Python 3 with PyTorch V1.10.0. For our approach, 64-dimensional word embeddings are used for code edits tokens, AST-diff tokens and comment tokens. The hidden states of the Bi-GRUs (encoder) and the GRU (decoder) in our model are 64 and 128 dimensions respectively. All GRUs have two layers.

In our model, \emph{Code Edit Encoder}, \emph{Syntax Change Encoder}, \emph{Old Comment Encoder} and the decoder are jointly trained to minimize the cross-entropy. During the training phase, we optimized the parameters of our model using Adam \cite{kingma2014adam} with a batch size of 32. We set the learning rate of Adam to 0.001. A dropout \cite{srivastava2014dropout} of 0.6 is used for dense layers before computing the final probability. The model with the best (smallest) validation perplexity is used for evaluation. A beam search of width 5 is used to generate the target sequence when testing.

\subsection{Evaluation Metrics}
We use Accuracy, Recall@5, METEOR \cite{banerjee2005meteor}, SARI \cite{xu2016optimizing}, GLUE \cite{napoles2015ground} and two metrics proposed by the authors of CUP \cite{liu2020automating} for this task, namely Average Edit Distance (AED) and Relative Edit Distance (RED), to evaluate our approach and the baselines.

Our evaluation metrics are defined as follows:
% \begin{itemize}[leftmargin=20pt]
\begin{itemize}
\item \textbf{Accuracy}: Accuracy represents the proportion of the test samples where \emph{correct comments} are generated at Top-1 among the total number of cases examined. Here, \emph{correct comments} refer to those that are identical to the ground-truth (i.e., written by developers).
\item \textbf{Recall@5}: Similar to Accuracy, Recall@5 is the proportion of the test samples where \emph{correct comments} are generated at Top-5.
\item \textbf{AED}: AED measures the average word-level edit distance required to change the predicted results from CUP into the ground-truth. This value indicates the distance between the generated comments and the ground truth: the smaller, the better. The AED metric is defined as follows:
\begin{equation}
    A E D=\frac{1}{N} \sum_{k=1}^{N} \text { edit\_distance }\left(\hat{\boldsymbol{y}}^{(k)}, \boldsymbol{y}^{(k)}\right)
\end{equation}
Where $N$ is the number of test samples, $edit\_distance$ is the word-level Levenshtein distance and $\hat{\boldsymbol{y}}^{(k)}$ refers to the comment generated for the $k_{th}$ sample.
\item \textbf{RED}: RED is similar to AED, but measures the average of relative edit distances. The RED metric is defined as follows:
\begin{equation}
    R E D=\frac{1}{N} \sum_{k=1}^{N} \frac{\text { edit\_distance }\left(\hat{\boldsymbol{y}}^{(k)}, \boldsymbol{y}^{(k)}\right)}{\text { edit\_distance }\left(\boldsymbol{x}^{(k)}, \boldsymbol{y}^{(k)}\right)}
\end{equation}
Where $\boldsymbol{x}^{(k)}$ is the old comment for the $k_{th}$ sample. If an approach’s RED is less than 1, and developers can expect to spend less effort updating comments by using this approach.

\item \textbf{METEOR}: METEOR (Metric for Evaluation of Translation with Explicit ORdering) is a metric for the evaluation of machine-translation output. The metric was designed to fix some of the problems found in the more popular BLEU metric and produce a good correlation with human judgment at the sentence or segment level. 
\item \textbf{SARI}: SARI is a metric used initially for evaluating automatic text simplification systems. The metric compares the predicted simplified sentences against the reference and the source sentences. It explicitly measures the goodness of words added, deleted and kept by the system.
\item \textbf{GLEU}: GLEU metric is a variant of BLEU proposed for evaluating grammatical error corrections using n-gram overlap with a set of reference sentences, as opposed to precision or recall of specific annotated errors.

\end{itemize}

\subsection{Result Analysis}

\begin{table}[t]
\caption{Comparisons of our approach with each baseline}
\label{tab:rq1_1}
% \resizebox{\linewidth}{!}
{%
\begin{tabular}{lllll}
\hline
\textbf{Approach}    & \textbf{Accuracy} & \textbf{Recall@5} & \textbf{AED}  & \textbf{RED}   \\ \hline
\textbf{CUP}         & 15.8\%            & 26.8\%            & 3.62          & 0.960          \\
\textbf{HEBCUP}      & \textbf{25.6\%}   & 27.6\%            & 3.52          & 0.896          \\
\textbf{HatCUP} & 24.3\%            & \textbf{35.2\%}   & \textbf{3.44} & \textbf{0.861} \\ \hline
\end{tabular}%
}
\end{table}

\begin{table}[t]
\caption{METEOR, SARI and GLUE scores}
\label{tab:rq1_2}
% \resizebox{\textwidth}{!}
\begin{threeparttable}          %这行要添加
{%
\begin{tabular}{llll}
\hline
\textbf{Approach}    & \textbf{METEOR} & \textbf{SARI}  & \textbf{GLUE} \\ \hline
\textbf{CUP}         & 51.22           & 38.62      & 50.30      \\
\textbf{HEBCUP}      & 53.96           & 41.29      & 54.14      \\
\textbf{HatCUP} & \textbf{58.52}  & \textbf{45.63} & \textbf{56.67}    \\ \hline
\end{tabular}%
}
 \begin{tablenotes}
    \footnotesize 
    \item * The scores are presented as percentage values between 0 and 100.
 \end{tablenotes}
\end{threeparttable}
\end{table}

\subsubsection{RQ1: The Effectiveness Evaluation}
To evaluate the effectiveness of our proposed model, HatCUP, we evaluate it and the baseline methods on the testing set in terms of various metrics. The evaluation results for the dataset are shown in Table \ref{tab:rq1_1} and Table \ref{tab:rq1_2}. From the tables, we can observe the following:
\begin{itemize}
\item HatCUP is slightly below HEBCUP by 5\% in terms of the accuracy metric, and there may be several reasons. First, the ground-truth is rather subjective as a modified comment by real developers, while the use of the accuracy metric in generation tasks is very demanding. It is not easy to guarantee that the modifications inferred by the model are consistent with the developer's, unless they are relatively simple modifications (i.e., modifying a single token). Second, HEBCUP is a heuristic-based approach specifically designed for this scenario, which pays attention to the changed code and performs token-level comment updates. While for CUP, which is also a deep learning method, HatCUP outperforms it in terms of accuracy by more than 50\%.
\item For the Recall@5 metric, HatCUP achieves the best results. We attribute this to the following reasons: First, it introduces the syntax change encoder and structure-guided attention mechanism compared to previous work, which utilizes code structure change hints. 
% Second, we introduced an attention matrix in the attention mechanism based on the relationship of AST nodes, capturing information outside of code changes. 
In addition, our newly developed \emph{edit or non-edit} mechanism can imitate the behavior of human developers to a large extent, which does not generate a full new comment but generate a series of edit actions to show how to update the old comment. So the comment will be updated as wishes.
\item HatCUP also outperforms the state-of-the-arts in terms of AED and RED. The AED metric drops from 3.52 to 3.44, which means for each comment, the developer can edit fewer words on average with HatCUP compared to the other tools. The lower RED metric also indicates that our approach can reduce the edits developers need to perform for just-in-time comment updating.
\item Our model is better at \emph{editing} comments, as shown by the results on METEOR, SARI, and GLEU in Table \ref{tab:rq1_2}. The three metrics are flexible in word order and are often used to evaluate comment generation methods in prior studies. 
\end{itemize}

In general, considerable improvements are achieved by HatCUP over CUP in terms of all metrics. Compared to HEBCUP, HatCUP performs much better on Recall@5 and outperforms it in AED and RED by substantial margins. This highlights that our approach can update comments more effectively and accurately than the baselines.

\begin{table*}[tb]
\caption{Impact of the key components}
\label{tab:rq2_1}
% \resizebox{\linewidth}{!}
{%
\begin{tabular}{llllllll}
\hline
\textbf{Approach}    & \textbf{Accuracy} & \textbf{Recall@5} & \textbf{AED}  & \textbf{RED}   & \textbf{METEOR} & \textbf{SARI}  & \textbf{GLUE} \\ \hline
\textbf{HatCUP-syntax} & 22.3\%  & 31.0\%  & 3.49 & 0.890 & 56.23 & 44.19 & 54.86 \\
\textbf{HatCUP-edit}     & 15.8\%  & 27.7\%  & 3.59 & 0.933 & 53.18 & 41.27 & 50.40 \\
\textbf{HatCUP} & \textbf{24.3\%}   & \textbf{35.2\%}   & \textbf{3.44} & \textbf{0.861} & \textbf{58.52}  & \textbf{45.63} & \textbf{56.67}    \\ \hline
\end{tabular}%
}
\end{table*}

\subsubsection{RQ2: The Effects of Key Components}
The key of our comment updating task is to effectively capture the relationship and references between code changes and comments. Previous deep learning-based work, i.e., CUP, has considered code changes and old comments. To better capture the potential code change information, we introduced a syntax change encoder together with a structure-guided attention mechanism. Additionally, we try to model the edit actions rather than generate comment sequences from scratch, defined as \emph{edit or non-edit} mechanism. Therefore, we want to determine if the two key components would improve the task of comment updating. To this end, we compare HatCUP with its two variants: 1) \textbf{HatCUP-syntax}, which does not use the \emph{Syntax Change encoder} and the structure-guided attention mechanism, and 2) \textbf{HatCUP-edit}, which removes the \emph{edit or non-edit} mechanism from HatCUP, generating the new comment directly instead of edit actions. The results are shown in Table \ref{tab:rq2_1}. It can be seen that:
\begin{itemize}
\item HatCUP performs better than two variants in terms of all metrics. For accuracy, the improvements achieved by HatCUP range from 2.0\% to 8.5\%; for Recall@5, HatCUP improves by at least 4.2\%, which means HatCUP can generate more correct comments than the variants. For AED and RED, HatCUP still achieves the lowest result. HatCUP minimizes the number of editing operations required for developers to update the old comments.
\item HatCUP-edit achieves the worst performance. We manually inspected the test results to determine why the performance declined so much. Based on our inspection, we find that HatCUP-edit model tends to generate the same comments as the old comments. Since the old comment and the new comment are closely related, training a model to directly generate a new comment risks having it learn to just copy the old one.
\item The introduction of the \emph{Syntax Change Encoder} and the structure-guided attention mechanism improves the effectiveness of the model to a certain extent. From another point of view, even though HatCUP-syntax does not obtain the top results as HatCUP, it still achieves considerable performance (e.g., with recall@5 over 30\%), which further confirms the strength of our approach. Moreover, since HatCUP-syntax does not rely on the AST information, it can be used as a light variant of our approach, which can help developers update comments just-in-time with incomplete code snippets.
\end{itemize}

In summary, HatCUP significantly outperforms the two variants. These results demonstrate that the structure information and the \emph{edit or non-edit} mechanism are helpful and valuable for this task.

\begin{figure}[t]
  \centering
  \includegraphics[width=\linewidth]{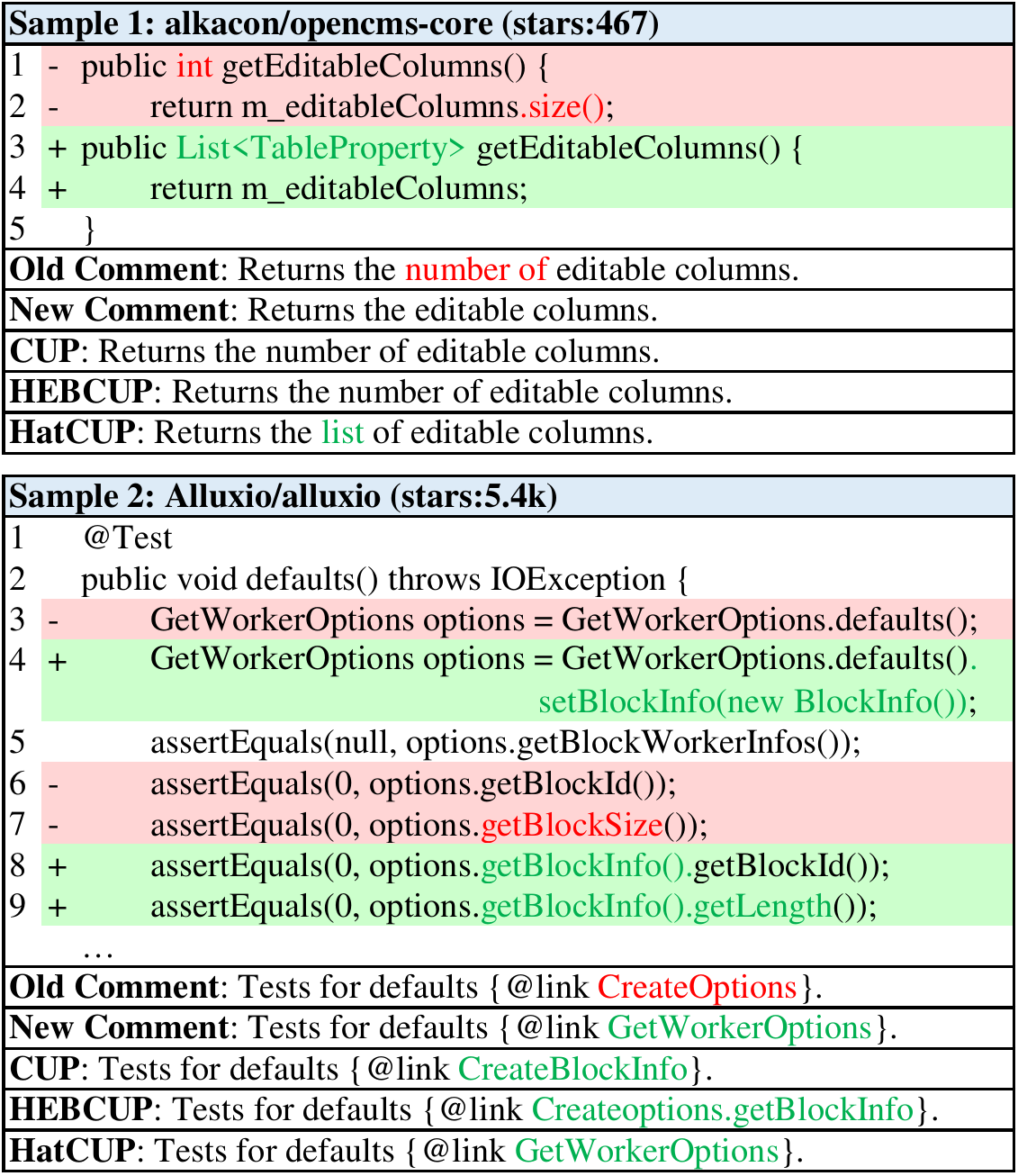}
  \caption{Case study}
  \label{fig:better_examples}
  \vspace{-1em}  % 调整与下文的间距
\end{figure}

\begin{table}[t]
\caption{Effectiveness in complex scenarios}
\label{tab:rq3_1}
\begin{tabular}{lllll}
\hline
\textbf{Approach}    & \textbf{Accuracy} & \textbf{Recall@5} & \textbf{AED} & \textbf{RED} \\ \hline
\textbf{CUP}    & 3.2\%   & 8.8\%    & 4.36   & 0.978     \\
\textbf{HEBCUP} & 0.0\%   & 1.1\%    & 4.41   & 0.992     \\
\textbf{HatCUP} & \textbf{7.8\%}   & \textbf{20.5\%}   & \textbf{4.27}   & \textbf{0.963}     \\ \hline
\end{tabular}
\vspace{-1em}  % 调整与下文的间距
\end{table}

\subsubsection{RQ3: The Effectiveness in Complex Scenarios}
To cope with comment updating in complex scenarios, three encoders are used and various attention mechanisms are introduced. At the same time, we wanted to check whether the \emph{edit or non-edit} mechanism could handle more complex updates. As illustrated by the authors, HEBCUP, can only work on code-indicative updates. It is difficult to manually define templates for comments whose updated contents do not appear within the code change content. Therefore, it is reasonable to assume that the instances HEBCUP cannot handle correctly are more complex. We isolated instances from the test dataset that HEBCUP could not handle, yielding a total of 6 844 samples. We evaluated HatCUP and the baselines on these samples. The results are shown in Table \ref{tab:rq3_1}. HatCUP performs much better on those complex instances when compared to the baselines. 

To better understand these performance differences, we manually inspect the test results of two examples, which have multiple code changes, including <INSERT>, <DEL>, and <UPDATE>. Existing approaches cannot capture any change based only on text analysis, while with multiple considerations, HatCUP can deal with these changes successfully. Considering sample 1 in Figure \ref{fig:better_examples}, the code snippet changed in the type of method and the return value, and the developer only deleted the span "\emph{number of}" in the comment. It is noticeable that there is no overlap between the changed tokens in the code snippet and those in the old comment. That is why the two state-of-the-art techniques cannot generate the correct comment (do not update the old comment). However, our model recognized that the token \emph{number} is not suitable for the new code due to the update of the method type. Then, it updates the token to \emph{list} according to the new type \emph{List<...>}, which is more suitable for the changed new code. For sample 2, the developer modified one method call, resulting in a change to the assignment of the variable \emph{option}. The variable \emph{option} is an instance of the class \emph{GetWorkerOptions}. The developer's intention was to test \emph{GetWorkerOptions}. We could not have given the correct suggestion without using AST to establish associations among the individual variable nodes.

In conclusion, different from the two baselines, which use text analysis to capture change information, HatCUP takes structure code changes into account, which covers the shortage of baselines. In addition, updating comments through an edit mechanism rather than writing new comments from scratch also makes sense.

\section{Discussion}
\label{sec:discussion}
In this section, we discuss the situations where HatCUP may fail, and the threats to the validity of this work.

\subsection{Where Does Our Approach Fail}
Although our method has proven superior to existing methods, there are still scenarios where HatCUP does not perform perfectly.

A common bad situation is that the code changes are too massive for HatCUP to handle perfectly. For example, the developer may rewrite the entire method, including the method name. In this case, the modified code can no longer be considered a variant of the original method. There is no connection between the old comment and the new code snippet. It is difficult for the model to update the old comment correctly by applying some edit actions.

Another situation, which we believe cannot simply be called a failure, is the optimization of language expression. For instance, the motivating example in Figure \ref{fig:motivation}, whose comment was updated with a conditional clause \emph{"... if text message was null"}, is a typical example of this situation. HatCUP may correctly update some of the comment phrases but not always all, which leads to inconsistency. 
% The decoder always tends to generate shorter sequences to decode the input state, i.e., with fewer editing actions to update the old comment to a new comment that matches the edited code. Therefore, our model will tend to generate simple phrases \emph{"or null"} added to the old comment. Although there is a gap between the updated annotation of the model and the real annotation, it does not convey the wrong meaning. We thus believe that our approach remains promising.

\subsection{Threats to Validity}
\subsubsection{External Validity}
A threat to external validity is related to GumTreeDiff we used to obtain the difference between ASTs. There is no guarantee that the syntax information extracted by GumTreeDiff is exactly correct. However, we manually checked 100 samples in the dataset and found only one incorrect mapping. Therefore, we believe the threat is limited.

\subsubsection{Internal Validity}
A threat to internal validity is related to the dataset we used. For comparison with CUP and HEBCUP, we directly used the dataset provided by them. The dataset is built only from Java projects and only contains updates of method comments, which may not be representative of all programming languages and comment types. 
% Considering that our approach is language-independent, we argue that our approach can be easily adapted to other programming languages.
Another threat is that the performance of HatCUP in solving complex cases is still not perfect. It needs more precise program analysis and efficient NLP algorithms.

\section{RELATED WORK}
\label{sec:relatedwork}
In this section, we discuss related work concerning code-comment inconsistency detection, comment updating and comment generation. All these works focus on maintaining comments, but have different emphases.

\subsection{Code-Comment Inconsistency Detection}
A large amount of work has been conducted by researchers to detect inconsistent comments. Most prior works targeted comments related to specific code properties \cite{tan2007icomment,tan2007hotcomments,tan2011acomment}. For instance, Tan et al. \cite{tan2007icomment,tan2007hotcomments,tan2011acomment} proposed several approaches to detect the consistency between code and comment concerning specific code properties, such as lock mechanisms \cite{tan2007icomment,tan2007hotcomments}, function calls \cite{tan2007icomment} and interrupts \cite{tan2011acomment}. They use static program analysis to check whether the source code conforms to specific rules. 
Some work has focused on specific types of comments \cite{huang2018identifying,sridhara2016automatically,gao2021automating}. For example, Huang et al. \cite{huang2018identifying} used the text mining-based methods to predict whether a comment contains self-admitted technical debt (SATD) (e.g., TODO, FIXME, HACK). 
%Rungroj et al. \cite{maipradit2020wait} first introduced the concept of “on-hold” SATD and proposed a tool \cite{maipradit2020automated} to automatically identify and remove the “on-hold” SATD. 
Sridhara \cite{sridhara2016automatically} proposed a technique to identify obsolete TODO comments based on information retrieval, linguistics and semantics. Gao et al. \cite{gao2021automating} proposed a deep learning-based approach TDCleaner, which outperforms Sridhara's by a large margin. Several studies focused on general comments. Ratol et al. \cite{ratol2017detecting} designed a rule-based approach named Fraco to detect fragile comments during identifier renaming. Panthaplackel et al. \cite{panthaplackel2021deep} developed a deep learning-based approach for just-in-time code-comment inconsistency detection by learning to relate comments and code changes.

\subsection{Comment Updating}
Following the work of code-comment inconsistency detection, some approaches have been proposed to focus on automatic comment updating. Liu et al. \cite{liu2020automating} are the first to propose a just-in-time comment updating technique, called CUP. The core idea of CUP is to leverage a neural sequence-to-sequence model to learn comment update patterns from old comments and changed code tokens; then, it can update the comment in time after the developer modifies the code. Lin et al. \cite{lin2021automated} performed an in-depth analysis on the effectiveness of CUP. They found that most of the successful updating conducted by CUP was related to a single token change. Therefore, for the case of single token modification in the code, they proposed HEBCUP, a heuristic-based approach, which achieves better performance on CUP. However, HEBCUP is not sufficiently effective beyond simple updates due to the limitation of heuristic rules.

\subsection{Comment Generation}
Source code comment generation has been studied by many researchers previously. In earlier studies, scholars tend to use template-based approaches \cite{haiduc2010supporting,haiduc2010use,eddy2013evaluating}. However, a well-designed template requires expert domain knowledge, which is not easy work. Consequently, IR-based approaches \cite{sridhara2010towards,wong2015clocom} have been proposed. To generate comments for Java methods, Sridhara et al. \cite{sridhara2010towards} use summary information in source code and manually define templates. ColCom \cite{wong2015clocom} proposed an approach that generates comments by reusing and tailoring comments of similar code snippets from open source projects. However, the retrieved comments may not correctly describe the semantics and behavior of code snippets, leading to the mismatches between code and comments. Recently, Neural Machine Translation (NMT) based models have been exploited to generate summaries for code snippets. CodeNN \cite{iyer2016summarizing} is an early attempt that uses only code token sequences, followed by various approaches that utilize AST \cite{hu2018deep,hu2020deepAST,leclair2019neural,alon2018code2seq,leclair2020improved,lin2021Improving}, API knowledge \cite{hu2018summarizing}, type information \cite{cai2020tag}, global context \cite{lin2021Improving,haque2020improved}, reinforcement learning \cite{wan2018improving,wang2020reinforcement}, multitask and dual learning \cite{wei2019code,xie2021exploiting,ye2020leveraging}, and pretrained language models \cite{feng2020codebert}.

\section{Conclusion}
\label{sec:conclusion}
We propose a new approach, HatCUP, for just-in-time comment updating. To the best of our knowledge, this is the first work that considers the code structure change information. Combined with code change graph analysis and data flow dependency analysis, we introduce a syntax change encoder together with a structure-guided attention mechanism to more fully utilize code structure change hints. Additionally, the \emph{edit or non-edit} mechanism, which is aimed at generating a sequence of edit actions to mimic human editing behavior, has proven better suited to the comment updating task than traditional approaches. Our results demonstrate that HatCUP outperforms the two state-of-the-art techniques and can substantially reduce developers’ efforts in updating comments.

% In the future, we plan to try other advanced techniques to further improve the performance of HatCUP, especially in complex scenarios. In addition, it would be meaningful to expand the dataset for comment updating task to incorporate more programming languages, which will benefit researchers interested in this study.

\begin{acks}
We thank the anonymous reviewers for their constructive comments. This research was supported, in part by NSFC 61832009, Cooperation Fund of Huawei-Nanjing University Next Generation Programming Innovation Lab (No. YBN2019105178SW27, No. YBN2019105178SW32). Any opinions, findings, and conclusions in this paper are those of the authors only and do not necessarily reflect the views of our sponsors.
\end{acks}

\bibliographystyle{ACM-Reference-Format}
\bibliography{main}


\begin{thebibliography}{51}

%%% the Web address.

\ifx \showCODEN    \undefined \def \showCODEN     #1{\unskip}     \fi
\ifx \showDOI      \undefined \def \showDOI       #1{#1}\fi
\ifx \showISBNx    \undefined \def \showISBNx     #1{\unskip}     \fi
\ifx \showISBNxiii \undefined \def \showISBNxiii  #1{\unskip}     \fi
\ifx \showISSN     \undefined \def \showISSN      #1{\unskip}     \fi
\ifx \showLCCN     \undefined \def \showLCCN      #1{\unskip}     \fi
\ifx \shownote     \undefined \def \shownote      #1{#1}          \fi
\ifx \showarticletitle \undefined \def \showarticletitle #1{#1}   \fi
\ifx \showURL      \undefined \def \showURL       {\relax}        \fi
% invisible to TeX
\providecommand\bibfield[2]{#2}
\providecommand\bibinfo[2]{#2}
\providecommand\natexlab[1]{#1}
\providecommand\showeprint[2][]{arXiv:#2}

\bibitem[\protect\citeauthoryear{Ahmad, Chakraborty, Ray, and Chang}{Ahmad
  et~al\mbox{.}}{2020}]%
        {ahmad2020transformer}
\bibfield{author}{\bibinfo{person}{Wasi Ahmad}, \bibinfo{person}{Saikat
  Chakraborty}, \bibinfo{person}{Baishakhi Ray}, {and} \bibinfo{person}{Kai-Wei
  Chang}.} \bibinfo{year}{2020}\natexlab{}.
\newblock \showarticletitle{A Transformer-based Approach for Source Code
  Summarization}. In \bibinfo{booktitle}{\emph{Proceedings of the 58th Annual
  Meeting of the Association for Computational Linguistics}}.
  \bibinfo{publisher}{Association for Computational Linguistics},
  \bibinfo{address}{Online}, \bibinfo{pages}{4998--5007}.
\newblock
\urldef\tempurl%
\url{https://doi.org/10.18653/v1/2020.acl-main.449}
\showDOI{\tempurl}


\bibitem[\protect\citeauthoryear{Alon, Brody, Levy, and Yahav}{Alon
  et~al\mbox{.}}{2018}]%
        {alon2018code2seq}
\bibfield{author}{\bibinfo{person}{Uri Alon}, \bibinfo{person}{Shaked Brody},
  \bibinfo{person}{Omer Levy}, {and} \bibinfo{person}{Eran Yahav}.}
  \bibinfo{year}{2018}\natexlab{}.
\newblock \showarticletitle{code2seq: Generating sequences from structured
  representations of code}.
\newblock \bibinfo{journal}{\emph{arXiv preprint arXiv:1808.01400}}
  (\bibinfo{year}{2018}).
\newblock


\bibitem[\protect\citeauthoryear{Banerjee and Lavie}{Banerjee and
  Lavie}{2005}]%
        {banerjee2005meteor}
\bibfield{author}{\bibinfo{person}{Satanjeev Banerjee} {and}
  \bibinfo{person}{Alon Lavie}.} \bibinfo{year}{2005}\natexlab{}.
\newblock \showarticletitle{METEOR: An automatic metric for MT evaluation with
  improved correlation with human judgments}. In
  \bibinfo{booktitle}{\emph{Proceedings of the acl workshop on intrinsic and
  extrinsic evaluation measures for machine translation and/or summarization}}.
  \bibinfo{pages}{65--72}.
\newblock


\bibitem[\protect\citeauthoryear{Cai, Liang, Xu, Li, Hao, and Chen}{Cai
  et~al\mbox{.}}{2020}]%
        {cai2020tag}
\bibfield{author}{\bibinfo{person}{Ruichu Cai}, \bibinfo{person}{Zhihao Liang},
  \bibinfo{person}{Boyan Xu}, \bibinfo{person}{Zijian Li},
  \bibinfo{person}{Yuexing Hao}, {and} \bibinfo{person}{Yao Chen}.}
  \bibinfo{year}{2020}\natexlab{}.
\newblock \showarticletitle{TAG: Type Auxiliary Guiding for Code Comment
  Generation}. In \bibinfo{booktitle}{\emph{Proceedings of the 58th Annual
  Meeting of the Association for Computational Linguistics}}.
  \bibinfo{pages}{291--301}.
\newblock


\bibitem[\protect\citeauthoryear{de~Souza, Anquetil, and de~Oliveira}{de~Souza
  et~al\mbox{.}}{2005}]%
        {de2005study}
\bibfield{author}{\bibinfo{person}{Sergio Cozzetti~B. de Souza},
  \bibinfo{person}{Nicolas Anquetil}, {and} \bibinfo{person}{K\'{a}thia~M. de
  Oliveira}.} \bibinfo{year}{2005}\natexlab{}.
\newblock \showarticletitle{A Study of the Documentation Essential to Software
  Maintenance}. In \bibinfo{booktitle}{\emph{Proceedings of the 23rd Annual
  International Conference on Design of Communication: Documenting \& Designing
  for Pervasive Information}} (Coventry, United Kingdom)
  \emph{(\bibinfo{series}{SIGDOC '05})}. \bibinfo{publisher}{Association for
  Computing Machinery}, \bibinfo{address}{New York, NY, USA},
  \bibinfo{pages}{68–75}.
\newblock
\showISBNx{1595931759}
\urldef\tempurl%
\url{https://doi.org/10.1145/1085313.1085331}
\showDOI{\tempurl}


\bibitem[\protect\citeauthoryear{Eddy, Robinson, Kraft, and Carver}{Eddy
  et~al\mbox{.}}{2013}]%
        {eddy2013evaluating}
\bibfield{author}{\bibinfo{person}{Brian~P Eddy}, \bibinfo{person}{Jeffrey~A
  Robinson}, \bibinfo{person}{Nicholas~A Kraft}, {and}
  \bibinfo{person}{Jeffrey~C Carver}.} \bibinfo{year}{2013}\natexlab{}.
\newblock \showarticletitle{Evaluating source code summarization techniques:
  Replication and expansion}. In \bibinfo{booktitle}{\emph{2013 21st
  International Conference on Program Comprehension (ICPC)}}. IEEE,
  \bibinfo{pages}{13--22}.
\newblock


\bibitem[\protect\citeauthoryear{Falleri, Morandat, Blanc, Martinez, and
  Monperrus}{Falleri et~al\mbox{.}}{2014}]%
        {falleri2014fine}
\bibfield{author}{\bibinfo{person}{Jean-R\'{e}my Falleri},
  \bibinfo{person}{Flor\'{e}al Morandat}, \bibinfo{person}{Xavier Blanc},
  \bibinfo{person}{Matias Martinez}, {and} \bibinfo{person}{Martin Monperrus}.}
  \bibinfo{year}{2014}\natexlab{}.
\newblock \showarticletitle{Fine-Grained and Accurate Source Code
  Differencing}. In \bibinfo{booktitle}{\emph{Proceedings of the 29th ACM/IEEE
  International Conference on Automated Software Engineering}} (Vasteras,
  Sweden) \emph{(\bibinfo{series}{ASE '14})}. \bibinfo{publisher}{Association
  for Computing Machinery}, \bibinfo{address}{New York, NY, USA},
  \bibinfo{pages}{313–324}.
\newblock
\showISBNx{9781450330138}
\urldef\tempurl%
\url{https://doi.org/10.1145/2642937.2642982}
\showDOI{\tempurl}


\bibitem[\protect\citeauthoryear{Feng, Guo, Tang, Duan, Feng, Gong, Shou, Qin,
  Liu, Jiang, et~al\mbox{.}}{Feng et~al\mbox{.}}{2020}]%
        {feng2020codebert}
\bibfield{author}{\bibinfo{person}{Zhangyin Feng}, \bibinfo{person}{Daya Guo},
  \bibinfo{person}{Duyu Tang}, \bibinfo{person}{Nan Duan},
  \bibinfo{person}{Xiaocheng Feng}, \bibinfo{person}{Ming Gong},
  \bibinfo{person}{Linjun Shou}, \bibinfo{person}{Bing Qin},
  \bibinfo{person}{Ting Liu}, \bibinfo{person}{Daxin Jiang}, {et~al\mbox{.}}}
  \bibinfo{year}{2020}\natexlab{}.
\newblock \showarticletitle{CodeBERT: A Pre-Trained Model for Programming and
  Natural Languages}. In \bibinfo{booktitle}{\emph{Proceedings of the 2020
  Conference on Empirical Methods in Natural Language Processing: Findings}}.
  \bibinfo{pages}{1536--1547}.
\newblock


\bibitem[\protect\citeauthoryear{Gao, Xia, Lo, Grundy, and Zimmermann}{Gao
  et~al\mbox{.}}{2021}]%
        {gao2021automating}
\bibfield{author}{\bibinfo{person}{Zhipeng Gao}, \bibinfo{person}{Xin Xia},
  \bibinfo{person}{David Lo}, \bibinfo{person}{John Grundy}, {and}
  \bibinfo{person}{Thomas Zimmermann}.} \bibinfo{year}{2021}\natexlab{}.
\newblock \showarticletitle{Automating the removal of obsolete TODO comments}.
  In \bibinfo{booktitle}{\emph{Proceedings of the 29th ACM Joint Meeting on
  European Software Engineering Conference and Symposium on the Foundations of
  Software Engineering}}. \bibinfo{pages}{218--229}.
\newblock


\bibitem[\protect\citeauthoryear{Guo, Ren, Lu, Feng, Tang, LIU, Zhou, Duan,
  Svyatkovskiy, Fu, Tufano, Deng, Clement, Drain, Sundaresan, Yin, Jiang, and
  Zhou}{Guo et~al\mbox{.}}{2021}]%
        {guo2021graphcodebert}
\bibfield{author}{\bibinfo{person}{Daya Guo}, \bibinfo{person}{Shuo Ren},
  \bibinfo{person}{Shuai Lu}, \bibinfo{person}{Zhangyin Feng},
  \bibinfo{person}{Duyu Tang}, \bibinfo{person}{Shujie LIU},
  \bibinfo{person}{Long Zhou}, \bibinfo{person}{Nan Duan},
  \bibinfo{person}{Alexey Svyatkovskiy}, \bibinfo{person}{Shengyu Fu},
  \bibinfo{person}{Michele Tufano}, \bibinfo{person}{Shao~Kun Deng},
  \bibinfo{person}{Colin Clement}, \bibinfo{person}{Dawn Drain},
  \bibinfo{person}{Neel Sundaresan}, \bibinfo{person}{Jian Yin},
  \bibinfo{person}{Daxin Jiang}, {and} \bibinfo{person}{Ming Zhou}.}
  \bibinfo{year}{2021}\natexlab{}.
\newblock \showarticletitle{GraphCode{\{}BERT{\}}: Pre-training Code
  Representations with Data Flow}. In \bibinfo{booktitle}{\emph{International
  Conference on Learning Representations}}.
\newblock
\urldef\tempurl%
\url{https://openreview.net/forum?id=jLoC4ez43PZ}
\showURL{%
\tempurl}


\bibitem[\protect\citeauthoryear{Haiduc, Aponte, and Marcus}{Haiduc
  et~al\mbox{.}}{2010a}]%
        {haiduc2010supporting}
\bibfield{author}{\bibinfo{person}{Sonia Haiduc}, \bibinfo{person}{Jairo
  Aponte}, {and} \bibinfo{person}{Andrian Marcus}.}
  \bibinfo{year}{2010}\natexlab{a}.
\newblock \showarticletitle{Supporting program comprehension with source code
  summarization}. In \bibinfo{booktitle}{\emph{2010 acm/ieee 32nd international
  conference on software engineering}}, Vol.~\bibinfo{volume}{2}. IEEE,
  \bibinfo{pages}{223--226}.
\newblock


\bibitem[\protect\citeauthoryear{Haiduc, Aponte, Moreno, and Marcus}{Haiduc
  et~al\mbox{.}}{2010b}]%
        {haiduc2010use}
\bibfield{author}{\bibinfo{person}{Sonia Haiduc}, \bibinfo{person}{Jairo
  Aponte}, \bibinfo{person}{Laura Moreno}, {and} \bibinfo{person}{Andrian
  Marcus}.} \bibinfo{year}{2010}\natexlab{b}.
\newblock \showarticletitle{On the use of automated text summarization
  techniques for summarizing source code}. In \bibinfo{booktitle}{\emph{2010
  17th Working Conference on Reverse Engineering}}. IEEE,
  \bibinfo{pages}{35--44}.
\newblock


\bibitem[\protect\citeauthoryear{Haque, LeClair, Wu, and McMillan}{Haque
  et~al\mbox{.}}{2020}]%
        {haque2020improved}
\bibfield{author}{\bibinfo{person}{Sakib Haque}, \bibinfo{person}{Alexander
  LeClair}, \bibinfo{person}{Lingfei Wu}, {and} \bibinfo{person}{Collin
  McMillan}.} \bibinfo{year}{2020}\natexlab{}.
\newblock \showarticletitle{Improved automatic summarization of subroutines via
  attention to file context}. In \bibinfo{booktitle}{\emph{Proceedings of the
  17th International Conference on Mining Software Repositories}}.
  \bibinfo{pages}{300--310}.
\newblock


\bibitem[\protect\citeauthoryear{He, Xu, Zhang, Hao, Feng, and Xu}{He
  et~al\mbox{.}}{2021}]%
        {he2021pyart}
\bibfield{author}{\bibinfo{person}{Xincheng He}, \bibinfo{person}{Lei Xu},
  \bibinfo{person}{Xiangyu Zhang}, \bibinfo{person}{Rui Hao},
  \bibinfo{person}{Yang Feng}, {and} \bibinfo{person}{Baowen Xu}.}
  \bibinfo{year}{2021}\natexlab{}.
\newblock \showarticletitle{PyART: Python API Recommendation in Real-Time}. In
  \bibinfo{booktitle}{\emph{2021 IEEE/ACM 43rd International Conference on
  Software Engineering (ICSE)}}. \bibinfo{publisher}{IEEE},
  \bibinfo{pages}{1634--1645}.
\newblock
\urldef\tempurl%
\url{https://doi.org/10.1109/ICSE43902.2021.00145}
\showDOI{\tempurl}


\bibitem[\protect\citeauthoryear{Hu, Li, Xia, Lo, and Jin}{Hu
  et~al\mbox{.}}{2018a}]%
        {hu2018deep}
\bibfield{author}{\bibinfo{person}{Xing Hu}, \bibinfo{person}{Ge Li},
  \bibinfo{person}{Xin Xia}, \bibinfo{person}{David Lo}, {and}
  \bibinfo{person}{Zhi Jin}.} \bibinfo{year}{2018}\natexlab{a}.
\newblock \showarticletitle{Deep code comment generation}. In
  \bibinfo{booktitle}{\emph{2018 IEEE/ACM 26th International Conference on
  Program Comprehension (ICPC)}}. IEEE, \bibinfo{pages}{200--20010}.
\newblock


\bibitem[\protect\citeauthoryear{Hu, Li, Xia, Lo, and Jin}{Hu
  et~al\mbox{.}}{2020}]%
        {hu2020deepAST}
\bibfield{author}{\bibinfo{person}{Xing Hu}, \bibinfo{person}{Ge Li},
  \bibinfo{person}{Xin Xia}, \bibinfo{person}{David Lo}, {and}
  \bibinfo{person}{Zhi Jin}.} \bibinfo{year}{2020}\natexlab{}.
\newblock \showarticletitle{Deep code comment generation with hybrid lexical
  and syntactical information}.
\newblock \bibinfo{journal}{\emph{Empirical Software Engineering}}
  \bibinfo{volume}{25}, \bibinfo{number}{3} (\bibinfo{year}{2020}),
  \bibinfo{pages}{2179--2217}.
\newblock


\bibitem[\protect\citeauthoryear{Hu, Li, Xia, Lo, Lu, and Jin}{Hu
  et~al\mbox{.}}{2018b}]%
        {hu2018summarizing}
\bibfield{author}{\bibinfo{person}{Xing Hu}, \bibinfo{person}{Ge Li},
  \bibinfo{person}{Xin Xia}, \bibinfo{person}{David Lo}, \bibinfo{person}{Shuai
  Lu}, {and} \bibinfo{person}{Zhi Jin}.} \bibinfo{year}{2018}\natexlab{b}.
\newblock \showarticletitle{Summarizing source code with transferred API
  knowledge}. In \bibinfo{booktitle}{\emph{Proceedings of the 27th
  International Joint Conference on Artificial Intelligence}}.
  \bibinfo{pages}{2269--2275}.
\newblock


\bibitem[\protect\citeauthoryear{Huang, Shihab, Xia, Lo, and Li}{Huang
  et~al\mbox{.}}{2018}]%
        {huang2018identifying}
\bibfield{author}{\bibinfo{person}{Qiao Huang}, \bibinfo{person}{Emad Shihab},
  \bibinfo{person}{Xin Xia}, \bibinfo{person}{David Lo}, {and}
  \bibinfo{person}{Shanping Li}.} \bibinfo{year}{2018}\natexlab{}.
\newblock \showarticletitle{Identifying self-admitted technical debt in open
  source projects using text mining}.
\newblock \bibinfo{journal}{\emph{Empirical Software Engineering}}
  \bibinfo{volume}{23}, \bibinfo{number}{1} (\bibinfo{year}{2018}),
  \bibinfo{pages}{418--451}.
\newblock


\bibitem[\protect\citeauthoryear{Iyer, Konstas, Cheung, and Zettlemoyer}{Iyer
  et~al\mbox{.}}{2016}]%
        {iyer2016summarizing}
\bibfield{author}{\bibinfo{person}{Srinivasan Iyer}, \bibinfo{person}{Ioannis
  Konstas}, \bibinfo{person}{Alvin Cheung}, {and} \bibinfo{person}{Luke
  Zettlemoyer}.} \bibinfo{year}{2016}\natexlab{}.
\newblock \showarticletitle{Summarizing source code using a neural attention
  model}. In \bibinfo{booktitle}{\emph{Proceedings of the 54th Annual Meeting
  of the Association for Computational Linguistics (Volume 1: Long Papers)}}.
  \bibinfo{pages}{2073--2083}.
\newblock


\bibitem[\protect\citeauthoryear{Kingma and Ba}{Kingma and Ba}{2014}]%
        {kingma2014adam}
\bibfield{author}{\bibinfo{person}{Diederik~P Kingma} {and}
  \bibinfo{person}{Jimmy Ba}.} \bibinfo{year}{2014}\natexlab{}.
\newblock \showarticletitle{Adam: A method for stochastic optimization}.
\newblock \bibinfo{journal}{\emph{arXiv preprint arXiv:1412.6980}}
  (\bibinfo{year}{2014}).
\newblock


\bibitem[\protect\citeauthoryear{LeClair, Haque, Wu, and McMillan}{LeClair
  et~al\mbox{.}}{2020}]%
        {leclair2020improved}
\bibfield{author}{\bibinfo{person}{Alexander LeClair}, \bibinfo{person}{Sakib
  Haque}, \bibinfo{person}{Lingfei Wu}, {and} \bibinfo{person}{Collin
  McMillan}.} \bibinfo{year}{2020}\natexlab{}.
\newblock \showarticletitle{Improved code summarization via a graph neural
  network}. In \bibinfo{booktitle}{\emph{Proceedings of the 28th International
  Conference on Program Comprehension}}. \bibinfo{pages}{184--195}.
\newblock


\bibitem[\protect\citeauthoryear{LeClair, Jiang, and McMillan}{LeClair
  et~al\mbox{.}}{2019}]%
        {leclair2019neural}
\bibfield{author}{\bibinfo{person}{Alexander LeClair}, \bibinfo{person}{Siyuan
  Jiang}, {and} \bibinfo{person}{Collin McMillan}.}
  \bibinfo{year}{2019}\natexlab{}.
\newblock \showarticletitle{A neural model for generating natural language
  summaries of program subroutines}. In \bibinfo{booktitle}{\emph{2019 IEEE/ACM
  41st International Conference on Software Engineering (ICSE)}}. IEEE,
  \bibinfo{pages}{795--806}.
\newblock


\bibitem[\protect\citeauthoryear{Liang and Zhu}{Liang and Zhu}{2018}]%
        {liang2018automatic}
\bibfield{author}{\bibinfo{person}{Yuding Liang} {and} \bibinfo{person}{Kenny
  Zhu}.} \bibinfo{year}{2018}\natexlab{}.
\newblock \showarticletitle{Automatic generation of text descriptive comments
  for code blocks}. In \bibinfo{booktitle}{\emph{Proceedings of the AAAI
  Conference on Artificial Intelligence}}, Vol.~\bibinfo{volume}{32}.
\newblock


\bibitem[\protect\citeauthoryear{Lin, Wang, Liu, Mao, and Bissyand{\'e}}{Lin
  et~al\mbox{.}}{2021b}]%
        {lin2021automated}
\bibfield{author}{\bibinfo{person}{Bo Lin}, \bibinfo{person}{Shangwen Wang},
  \bibinfo{person}{Kui Liu}, \bibinfo{person}{Xiaoguang Mao}, {and}
  \bibinfo{person}{Tegawend{\'e}~F Bissyand{\'e}}.}
  \bibinfo{year}{2021}\natexlab{b}.
\newblock \showarticletitle{Automated Comment Update: How Far are We?}. In
  \bibinfo{booktitle}{\emph{2021 IEEE/ACM 29th International Conference on
  Program Comprehension (ICPC)}}. IEEE, \bibinfo{pages}{36--46}.
\newblock
\urldef\tempurl%
\url{https://doi.org/10.1109/ICPC52881.2021.00013}
\showDOI{\tempurl}


\bibitem[\protect\citeauthoryear{Lin, Ouyang, Zhuang, Chen, Li, and Wu}{Lin
  et~al\mbox{.}}{2021a}]%
        {lin2021Improving}
\bibfield{author}{\bibinfo{person}{Chen Lin}, \bibinfo{person}{Zhichao Ouyang},
  \bibinfo{person}{Junqing Zhuang}, \bibinfo{person}{Jianqiang Chen},
  \bibinfo{person}{Hui Li}, {and} \bibinfo{person}{Rongxin Wu}.}
  \bibinfo{year}{2021}\natexlab{a}.
\newblock \showarticletitle{Improving code summarization with block-wise
  abstract syntax tree splitting}. In \bibinfo{booktitle}{\emph{2021 IEEE/ACM
  29th International Conference on Program Comprehension (ICPC)}}. IEEE,
  \bibinfo{pages}{184--195}.
\newblock
\urldef\tempurl%
\url{https://doi.org/10.1109/ICPC52881.2021.00026}
\showDOI{\tempurl}


\bibitem[\protect\citeauthoryear{Liu, Xia, Lo, Yan, and Li}{Liu
  et~al\mbox{.}}{2021}]%
        {liu2021just}
\bibfield{author}{\bibinfo{person}{Zhongxin Liu}, \bibinfo{person}{Xin Xia},
  \bibinfo{person}{David Lo}, \bibinfo{person}{Meng Yan}, {and}
  \bibinfo{person}{Shanping Li}.} \bibinfo{year}{2021}\natexlab{}.
\newblock \showarticletitle{Just-In-Time Obsolete Comment Detection and
  Update}.
\newblock \bibinfo{journal}{\emph{IEEE Transactions on Software Engineering}}
  (\bibinfo{year}{2021}), \bibinfo{pages}{1--1}.
\newblock
\urldef\tempurl%
\url{https://doi.org/10.1109/TSE.2021.3138909}
\showDOI{\tempurl}


\bibitem[\protect\citeauthoryear{Liu, Xia, Treude, Lo, and Li}{Liu
  et~al\mbox{.}}{2019}]%
        {liu2019automatic}
\bibfield{author}{\bibinfo{person}{Zhongxin Liu}, \bibinfo{person}{Xin Xia},
  \bibinfo{person}{Christoph Treude}, \bibinfo{person}{David Lo}, {and}
  \bibinfo{person}{Shanping Li}.} \bibinfo{year}{2019}\natexlab{}.
\newblock \showarticletitle{Automatic generation of pull request descriptions}.
  In \bibinfo{booktitle}{\emph{2019 34th IEEE/ACM International Conference on
  Automated Software Engineering (ASE)}}. IEEE, \bibinfo{pages}{176--188}.
\newblock


\bibitem[\protect\citeauthoryear{Liu, Xia, Yan, and Li}{Liu
  et~al\mbox{.}}{2020}]%
        {liu2020automating}
\bibfield{author}{\bibinfo{person}{Zhongxin Liu}, \bibinfo{person}{Xin Xia},
  \bibinfo{person}{Meng Yan}, {and} \bibinfo{person}{Shanping Li}.}
  \bibinfo{year}{2020}\natexlab{}.
\newblock \showarticletitle{Automating Just-in-Time Comment Updating}. In
  \bibinfo{booktitle}{\emph{Proceedings of the 35th IEEE/ACM International
  Conference on Automated Software Engineering}}.
  \bibinfo{publisher}{Association for Computing Machinery},
  \bibinfo{address}{New York, NY, USA}, \bibinfo{pages}{585–597}.
\newblock
\showISBNx{9781450367684}
\urldef\tempurl%
\url{https://doi.org/10.1145/3324884.3416581}
\showURL{%
\tempurl}


\bibitem[\protect\citeauthoryear{Napoles, Sakaguchi, Post, and
  Tetreault}{Napoles et~al\mbox{.}}{2015}]%
        {napoles2015ground}
\bibfield{author}{\bibinfo{person}{Courtney Napoles}, \bibinfo{person}{Keisuke
  Sakaguchi}, \bibinfo{person}{Matt Post}, {and} \bibinfo{person}{Joel
  Tetreault}.} \bibinfo{year}{2015}\natexlab{}.
\newblock \showarticletitle{Ground truth for grammatical error correction
  metrics}. In \bibinfo{booktitle}{\emph{Proceedings of the 53rd Annual Meeting
  of the Association for Computational Linguistics and the 7th International
  Joint Conference on Natural Language Processing (Volume 2: Short Papers)}}.
  \bibinfo{pages}{588--593}.
\newblock


\bibitem[\protect\citeauthoryear{Panthaplackel, Gligoric, Mooney, and
  Li}{Panthaplackel et~al\mbox{.}}{2020}]%
        {panthaplackel2020associating}
\bibfield{author}{\bibinfo{person}{Sheena Panthaplackel},
  \bibinfo{person}{Milos Gligoric}, \bibinfo{person}{Raymond~J. Mooney}, {and}
  \bibinfo{person}{Junyi~Jessy Li}.} \bibinfo{year}{2020}\natexlab{}.
\newblock \showarticletitle{Associating Natural Language Comment and Source
  Code Entities}.
\newblock \bibinfo{journal}{\emph{Proceedings of the AAAI Conference on
  Artificial Intelligence}} \bibinfo{volume}{34}, \bibinfo{number}{05}
  (\bibinfo{date}{Apr.} \bibinfo{year}{2020}), \bibinfo{pages}{8592--8599}.
\newblock
\urldef\tempurl%
\url{https://doi.org/10.1609/aaai.v34i05.6382}
\showDOI{\tempurl}


\bibitem[\protect\citeauthoryear{Panthaplackel, Li, Gligoric, and
  Mooney}{Panthaplackel et~al\mbox{.}}{2021}]%
        {panthaplackel2021deep}
\bibfield{author}{\bibinfo{person}{Sheena Panthaplackel},
  \bibinfo{person}{Junyi~Jessy Li}, \bibinfo{person}{Milos Gligoric}, {and}
  \bibinfo{person}{Raymond~J Mooney}.} \bibinfo{year}{2021}\natexlab{}.
\newblock \showarticletitle{Deep Just-In-Time Inconsistency Detection Between
  Comments and Source Code}. In \bibinfo{booktitle}{\emph{Proceedings of the
  AAAI Conference on Artificial Intelligence}}, Vol.~\bibinfo{volume}{35}.
  \bibinfo{pages}{427--435}.
\newblock


\bibitem[\protect\citeauthoryear{Ratol and Robillard}{Ratol and
  Robillard}{2017}]%
        {ratol2017detecting}
\bibfield{author}{\bibinfo{person}{Inderjot~Kaur Ratol} {and}
  \bibinfo{person}{Martin~P. Robillard}.} \bibinfo{year}{2017}\natexlab{}.
\newblock \showarticletitle{Detecting fragile comments}. In
  \bibinfo{booktitle}{\emph{2017 32nd IEEE/ACM International Conference on
  Automated Software Engineering (ASE)}}. \bibinfo{pages}{112--122}.
\newblock
\urldef\tempurl%
\url{https://doi.org/10.1109/ASE.2017.8115624}
\showDOI{\tempurl}


\bibitem[\protect\citeauthoryear{Shin, Polosukhin, and Song}{Shin
  et~al\mbox{.}}{2018}]%
        {shin2018towards}
\bibfield{author}{\bibinfo{person}{Richard Shin}, \bibinfo{person}{Illia
  Polosukhin}, {and} \bibinfo{person}{Dawn Song}.}
  \bibinfo{year}{2018}\natexlab{}.
\newblock \showarticletitle{Towards specification-directed program repair}. In
  \bibinfo{booktitle}{\emph{International Conference on Learning
  Representations Workshop}}.
\newblock
\urldef\tempurl%
\url{https://openreview.net/forum?id=B1iZRFkwz}
\showURL{%
\tempurl}


\bibitem[\protect\citeauthoryear{Sridhara}{Sridhara}{2016}]%
        {sridhara2016automatically}
\bibfield{author}{\bibinfo{person}{Giriprasad Sridhara}.}
  \bibinfo{year}{2016}\natexlab{}.
\newblock \showarticletitle{Automatically detecting the up-to-date status of
  ToDo comments in Java programs}. In \bibinfo{booktitle}{\emph{Proceedings of
  the 9th India Software Engineering Conference}}. \bibinfo{pages}{16--25}.
\newblock


\bibitem[\protect\citeauthoryear{Sridhara, Hill, Muppaneni, Pollock, and
  Vijay-Shanker}{Sridhara et~al\mbox{.}}{2010}]%
        {sridhara2010towards}
\bibfield{author}{\bibinfo{person}{Giriprasad Sridhara}, \bibinfo{person}{Emily
  Hill}, \bibinfo{person}{Divya Muppaneni}, \bibinfo{person}{Lori Pollock},
  {and} \bibinfo{person}{K. Vijay-Shanker}.} \bibinfo{year}{2010}\natexlab{}.
\newblock \showarticletitle{Towards Automatically Generating Summary Comments
  for Java Methods}. In \bibinfo{booktitle}{\emph{Proceedings of the IEEE/ACM
  International Conference on Automated Software Engineering}} (Antwerp,
  Belgium) \emph{(\bibinfo{series}{ASE '10})}. \bibinfo{publisher}{Association
  for Computing Machinery}, \bibinfo{address}{New York, NY, USA},
  \bibinfo{pages}{43–52}.
\newblock
\showISBNx{9781450301169}
\urldef\tempurl%
\url{https://doi.org/10.1145/1858996.1859006}
\showDOI{\tempurl}


\bibitem[\protect\citeauthoryear{Srivastava, Hinton, Krizhevsky, Sutskever, and
  Salakhutdinov}{Srivastava et~al\mbox{.}}{2014}]%
        {srivastava2014dropout}
\bibfield{author}{\bibinfo{person}{Nitish Srivastava},
  \bibinfo{person}{Geoffrey Hinton}, \bibinfo{person}{Alex Krizhevsky},
  \bibinfo{person}{Ilya Sutskever}, {and} \bibinfo{person}{Ruslan
  Salakhutdinov}.} \bibinfo{year}{2014}\natexlab{}.
\newblock \showarticletitle{Dropout: a simple way to prevent neural networks
  from overfitting}.
\newblock \bibinfo{journal}{\emph{The journal of machine learning research}}
  \bibinfo{volume}{15}, \bibinfo{number}{1} (\bibinfo{year}{2014}),
  \bibinfo{pages}{1929--1958}.
\newblock


\bibitem[\protect\citeauthoryear{Tan, Yuan, Krishna, and Zhou}{Tan
  et~al\mbox{.}}{2007b}]%
        {tan2007icomment}
\bibfield{author}{\bibinfo{person}{Lin Tan}, \bibinfo{person}{Ding Yuan},
  \bibinfo{person}{Gopal Krishna}, {and} \bibinfo{person}{Yuanyuan Zhou}.}
  \bibinfo{year}{2007}\natexlab{b}.
\newblock \showarticletitle{/*icomment: Bugs or Bad Comments?*/}. In
  \bibinfo{booktitle}{\emph{Proceedings of Twenty-First ACM SIGOPS Symposium on
  Operating Systems Principles}} (Stevenson, Washington, USA)
  \emph{(\bibinfo{series}{SOSP '07})}. \bibinfo{publisher}{Association for
  Computing Machinery}, \bibinfo{address}{New York, NY, USA},
  \bibinfo{pages}{145–158}.
\newblock
\showISBNx{9781595935915}
\urldef\tempurl%
\url{https://doi.org/10.1145/1294261.1294276}
\showDOI{\tempurl}


\bibitem[\protect\citeauthoryear{Tan, Yuan, and Zhou}{Tan
  et~al\mbox{.}}{2007a}]%
        {tan2007hotcomments}
\bibfield{author}{\bibinfo{person}{Lin Tan}, \bibinfo{person}{Ding Yuan}, {and}
  \bibinfo{person}{Yuanyuan Zhou}.} \bibinfo{year}{2007}\natexlab{a}.
\newblock \showarticletitle{Hotcomments: how to make program comments more
  useful?}. In \bibinfo{booktitle}{\emph{HotOS}}, Vol.~\bibinfo{volume}{7}.
  \bibinfo{pages}{49--54}.
\newblock


\bibitem[\protect\citeauthoryear{Tan, Zhou, and Padioleau}{Tan
  et~al\mbox{.}}{2011}]%
        {tan2011acomment}
\bibfield{author}{\bibinfo{person}{Lin Tan}, \bibinfo{person}{Yuanyuan Zhou},
  {and} \bibinfo{person}{Yoann Padioleau}.} \bibinfo{year}{2011}\natexlab{}.
\newblock \showarticletitle{aComment: mining annotations from comments and code
  to detect interrupt related concurrency bugs}. In
  \bibinfo{booktitle}{\emph{2011 33rd International Conference on Software
  Engineering (ICSE)}}. IEEE, \bibinfo{pages}{11--20}.
\newblock


\bibitem[\protect\citeauthoryear{Tufano, Pantiuchina, Watson, Bavota, and
  Poshyvanyk}{Tufano et~al\mbox{.}}{2019}]%
        {tufano2019learning}
\bibfield{author}{\bibinfo{person}{Michele Tufano}, \bibinfo{person}{Jevgenija
  Pantiuchina}, \bibinfo{person}{Cody Watson}, \bibinfo{person}{Gabriele
  Bavota}, {and} \bibinfo{person}{Denys Poshyvanyk}.}
  \bibinfo{year}{2019}\natexlab{}.
\newblock \showarticletitle{On learning meaningful code changes via neural
  machine translation}. In \bibinfo{booktitle}{\emph{2019 IEEE/ACM 41st
  International Conference on Software Engineering (ICSE)}}. IEEE,
  \bibinfo{pages}{25--36}.
\newblock


\bibitem[\protect\citeauthoryear{Vinyals, Fortunato, and Jaitly}{Vinyals
  et~al\mbox{.}}{2015}]%
        {vinyals2015pointer}
\bibfield{author}{\bibinfo{person}{Oriol Vinyals}, \bibinfo{person}{Meire
  Fortunato}, {and} \bibinfo{person}{Navdeep Jaitly}.}
  \bibinfo{year}{2015}\natexlab{}.
\newblock \showarticletitle{Pointer Networks}.
\newblock \bibinfo{journal}{\emph{Advances in Neural Information Processing
  Systems}}  \bibinfo{volume}{28} (\bibinfo{year}{2015}),
  \bibinfo{pages}{2692--2700}.
\newblock


\bibitem[\protect\citeauthoryear{Wan, Zhao, Yang, Xu, Ying, Wu, and Yu}{Wan
  et~al\mbox{.}}{2018}]%
        {wan2018improving}
\bibfield{author}{\bibinfo{person}{Yao Wan}, \bibinfo{person}{Zhou Zhao},
  \bibinfo{person}{Min Yang}, \bibinfo{person}{Guandong Xu},
  \bibinfo{person}{Haochao Ying}, \bibinfo{person}{Jian Wu}, {and}
  \bibinfo{person}{Philip~S. Yu}.} \bibinfo{year}{2018}\natexlab{}.
\newblock \bibinfo{booktitle}{\emph{Improving Automatic Source Code
  Summarization via Deep Reinforcement Learning}}.
\newblock \bibinfo{publisher}{Association for Computing Machinery},
  \bibinfo{address}{New York, NY, USA}, \bibinfo{pages}{397–407}.
\newblock
\showISBNx{9781450359375}
\urldef\tempurl%
\url{https://doi.org/10.1145/3238147.3238206}
\showURL{%
\tempurl}


\bibitem[\protect\citeauthoryear{Wang, Zhang, Sui, Wan, Zhao, Wu, Yu, and
  Xu}{Wang et~al\mbox{.}}{2020}]%
        {wang2020reinforcement}
\bibfield{author}{\bibinfo{person}{Wenhua Wang}, \bibinfo{person}{Yuqun Zhang},
  \bibinfo{person}{Yulei Sui}, \bibinfo{person}{Yao Wan}, \bibinfo{person}{Zhou
  Zhao}, \bibinfo{person}{Jian Wu}, \bibinfo{person}{Philip Yu}, {and}
  \bibinfo{person}{Guandong Xu}.} \bibinfo{year}{2020}\natexlab{}.
\newblock \showarticletitle{Reinforcement-learning-guided source code
  summarization via hierarchical attention}.
\newblock \bibinfo{journal}{\emph{IEEE Transactions on software Engineering}}
  (\bibinfo{year}{2020}).
\newblock


\bibitem[\protect\citeauthoryear{Wei, Li, Xia, Fu, and Jin}{Wei
  et~al\mbox{.}}{2019}]%
        {wei2019code}
\bibfield{author}{\bibinfo{person}{Bolin Wei}, \bibinfo{person}{Ge Li},
  \bibinfo{person}{Xin Xia}, \bibinfo{person}{Zhiyi Fu}, {and}
  \bibinfo{person}{Zhi Jin}.} \bibinfo{year}{2019}\natexlab{}.
\newblock \showarticletitle{Code Generation as a Dual Task of Code
  Summarization}.
\newblock \bibinfo{journal}{\emph{Advances in Neural Information Processing
  Systems}}  \bibinfo{volume}{32} (\bibinfo{year}{2019}),
  \bibinfo{pages}{6563--6573}.
\newblock


\bibitem[\protect\citeauthoryear{Wen, Nagy, Bavota, and Lanza}{Wen
  et~al\mbox{.}}{2019}]%
        {wen2019large}
\bibfield{author}{\bibinfo{person}{Fengcai Wen}, \bibinfo{person}{Csaba Nagy},
  \bibinfo{person}{Gabriele Bavota}, {and} \bibinfo{person}{Michele Lanza}.}
  \bibinfo{year}{2019}\natexlab{}.
\newblock \showarticletitle{A Large-Scale Empirical Study on Code-Comment
  Inconsistencies}. In \bibinfo{booktitle}{\emph{2019 IEEE/ACM 27th
  International Conference on Program Comprehension (ICPC)}}.
  \bibinfo{publisher}{IEEE}, \bibinfo{pages}{53--64}.
\newblock
\urldef\tempurl%
\url{https://doi.org/10.1109/ICPC.2019.00019}
\showDOI{\tempurl}


\bibitem[\protect\citeauthoryear{Wong, Liu, and Tan}{Wong
  et~al\mbox{.}}{2015}]%
        {wong2015clocom}
\bibfield{author}{\bibinfo{person}{Edmund Wong}, \bibinfo{person}{Taiyue Liu},
  {and} \bibinfo{person}{Lin Tan}.} \bibinfo{year}{2015}\natexlab{}.
\newblock \showarticletitle{Clocom: Mining existing source code for automatic
  comment generation}. In \bibinfo{booktitle}{\emph{2015 IEEE 22nd
  International Conference on Software Analysis, Evolution, and Reengineering
  (SANER)}}. IEEE, \bibinfo{pages}{380--389}.
\newblock


\bibitem[\protect\citeauthoryear{Xia, Bao, Lo, Xing, Hassan, and Li}{Xia
  et~al\mbox{.}}{2017}]%
        {xia2017measuring}
\bibfield{author}{\bibinfo{person}{Xin Xia}, \bibinfo{person}{Lingfeng Bao},
  \bibinfo{person}{David Lo}, \bibinfo{person}{Zhenchang Xing},
  \bibinfo{person}{Ahmed~E Hassan}, {and} \bibinfo{person}{Shanping Li}.}
  \bibinfo{year}{2017}\natexlab{}.
\newblock \showarticletitle{Measuring program comprehension: A large-scale
  field study with professionals}.
\newblock \bibinfo{journal}{\emph{IEEE Transactions on Software Engineering}}
  \bibinfo{volume}{44}, \bibinfo{number}{10} (\bibinfo{year}{2017}),
  \bibinfo{pages}{951--976}.
\newblock


\bibitem[\protect\citeauthoryear{Xie, Ye, Sun, and Zhang}{Xie
  et~al\mbox{.}}{2021}]%
        {xie2021exploiting}
\bibfield{author}{\bibinfo{person}{Rui Xie}, \bibinfo{person}{Wei Ye},
  \bibinfo{person}{Jinan Sun}, {and} \bibinfo{person}{Shikun Zhang}.}
  \bibinfo{year}{2021}\natexlab{}.
\newblock \showarticletitle{Exploiting Method Names to Improve Code
  Summarization: A Deliberation Multi-Task Learning Approach}. In
  \bibinfo{booktitle}{\emph{2021 IEEE/ACM 29th International Conference on
  Program Comprehension (ICPC)}}. IEEE.
\newblock


\bibitem[\protect\citeauthoryear{Xu, Napoles, Pavlick, Chen, and
  Callison-Burch}{Xu et~al\mbox{.}}{2016}]%
        {xu2016optimizing}
\bibfield{author}{\bibinfo{person}{Wei Xu}, \bibinfo{person}{Courtney Napoles},
  \bibinfo{person}{Ellie Pavlick}, \bibinfo{person}{Quanze Chen}, {and}
  \bibinfo{person}{Chris Callison-Burch}.} \bibinfo{year}{2016}\natexlab{}.
\newblock \showarticletitle{Optimizing statistical machine translation for text
  simplification}.
\newblock \bibinfo{journal}{\emph{Transactions of the Association for
  Computational Linguistics}}  \bibinfo{volume}{4} (\bibinfo{year}{2016}),
  \bibinfo{pages}{401--415}.
\newblock


\bibitem[\protect\citeauthoryear{Ye, Xie, Zhang, Hu, Wang, and Zhang}{Ye
  et~al\mbox{.}}{2020}]%
        {ye2020leveraging}
\bibfield{author}{\bibinfo{person}{Wei Ye}, \bibinfo{person}{Rui Xie},
  \bibinfo{person}{Jinglei Zhang}, \bibinfo{person}{Tianxiang Hu},
  \bibinfo{person}{Xiaoyin Wang}, {and} \bibinfo{person}{Shikun Zhang}.}
  \bibinfo{year}{2020}\natexlab{}.
\newblock \bibinfo{booktitle}{\emph{Leveraging Code Generation to Improve Code
  Retrieval and Summarization via Dual Learning}}.
\newblock \bibinfo{publisher}{Association for Computing Machinery},
  \bibinfo{address}{New York, NY, USA}, \bibinfo{pages}{2309–2319}.
\newblock
\showISBNx{9781450370233}
\urldef\tempurl%
\url{https://doi.org/10.1145/3366423.3380295}
\showURL{%
\tempurl}


\bibitem[\protect\citeauthoryear{Zhang, Wang, Zhang, Sun, and Liu}{Zhang
  et~al\mbox{.}}{2020}]%
        {zhang2020retrieval}
\bibfield{author}{\bibinfo{person}{Jian Zhang}, \bibinfo{person}{Xu Wang},
  \bibinfo{person}{Hongyu Zhang}, \bibinfo{person}{Hailong Sun}, {and}
  \bibinfo{person}{Xudong Liu}.} \bibinfo{year}{2020}\natexlab{}.
\newblock \showarticletitle{Retrieval-based neural source code summarization}.
  In \bibinfo{booktitle}{\emph{2020 IEEE/ACM 42nd International Conference on
  Software Engineering (ICSE)}}. IEEE, \bibinfo{pages}{1385--1397}.
\newblock


\end{thebibliography}

\appendix

\end{document}